\documentclass{lmcs} %
\pdfoutput=1

\usepackage{lastpage}
\lmcsdoi{22}{2}{24}
\lmcsheading{}{\pageref{LastPage}}{}{}%
{Dec.~20,~2024}{May~27,~2026}{}


\newcommand{\myenvalias}[2]{\newenvironment{#1}{\begin{#2}}{\end{#2}}}
\myenvalias{theorem}{thm}
\myenvalias{lemma}{lem}
\myenvalias{corollary}{cor}
\myenvalias{definition}{defi}
\myenvalias{notation}{nota}
\myenvalias{example}{exa}
\myenvalias{remark}{rem}

\usepackage{stmaryrd}
\usepackage{xcolor} %
\providecolor{Blue4}{RGB}{0, 0, 100}
\providecolor{Green4}{RGB}{0, 139, 0}
\usepackage{my_msc}

\usepackage{subfiles}
\usepackage[textsize=footnotesize, textwidth=25mm]{todonotes}
\usepackage{logic9-thm}
\usepackage{quotes}
\usepackage{subcaption}
\usepackage[hyperref,quotation]{knowledge} %
\usepackage{hyperref}
\usepackage[capitalise, noabbrev]{cleveref}
\usepackage{tikz}
\usepackage[utf8]{inputenc}

\usetikzlibrary{
   arrows.meta,bending,positioning,backgrounds,
   automata,arrows,decorations.pathreplacing,calligraphy,
   shapes.symbols
   }

\pgfdeclarelayer{background}
\pgfdeclarelayer{foreground}

\pgfsetlayers{background,main,foreground}

\usepackage{cancel}

\newcommand{\Ex}{\mathit{Tr}}

\newcommand{\mscn}{\texttt{msc}}
\newcommand{\Ch}{\mathit{Ch}}

\newcommand{\mbord}{\mathop{<_{\mb}}}

\newcommand{\msg}{\mathop{\mathrm{msg}}}

\newcommand{\act}[1]{\xrightarrow{#1}}

\newcommand{\ms}{\text{\texttt{ms}}}
\newcommand{\G}{H}

\newcommand{\mb}{\texttt{mb}}
\newcommand{\ptp}{\texttt{p2p}}

\newcommand{\Proc}{\mathbb{P}}
\newcommand{\Msg}{\mathbb{M}}
\newcommand{\Bufset}{\mathsf{B}}
\newcommand{\bufelt}{\mathsf{b}}
\newcommand{\Glob}{G}
\newcommand{\Msc}{\Mm}
\newcommand{\executable}[1]{\stackrel{#1}{\rightsquigarrow}}

\makeatletter
\newcommand{\dashbar}[2][\mathop]{#1{\mathpalette\df@over{{\dashfill}{#2}}}}
\newcommand{\df@over}[2]{\df@@over#1#2}
\newcommand\df@@over[3]{%
  \vbox{
    \offinterlineskip
    \ialign{##\cr
      #2{#1}\cr
      \noalign{\kern1pt}
      $\m@th#1#3$\cr
    }
  }%
}
\newcommand{\dashfill}[1]{%
  \kern-.5pt
  \xleaders\hbox{\kern.5pt\vrule height.6pt width \dash@width{#1}\kern.5pt}\hfill
  \kern-.5pt
}
\newcommand{\dash@width}[1]{%
  \ifx#1\displaystyle
    2pt
  \else
    \ifx#1\textstyle
      1.5pt
    \else
      \ifx#1\scriptstyle
        1.25pt
      \else
        \ifx#1\scriptscriptstyle
          1pt
        \fi
      \fi
    \fi
  \fi
}
\makeatother

\ifx\mscset\undefined
\else
  \mscset{
      /msc/draw head/.initial=none,       %
      /msc/draw head/.default=none,
      /msc/draw frame/.initial=none, 
      /msc/draw frame/.default=none,
      instance end/.initial=foot,
      /msc/left inline overlap=7.5pt,
      /msc/right inline overlap=7.5pt,
      stop width=0.2cm,
      message/.style={arrows=Circle->. Circle,shorten > = -2.5pt,shorten < = -2.5pt,thick},
      /msc/region width = 0pt,
      /msc/foot height = 0pt,
      /msc/foot distance = 2pt,
      /msc/left environment distance = 0pt,
      /msc/right environment distance = 0pt,
      /msc/head top distance = 0pt,
      /msc/title distance = 0pt,
      /msc/title top distance = 0pt
      }
  \tikzset{>=stealth}
  \newcommand{\unmatched}[4][]{\mess[arrows=
  {Circle->| Rectangle[white,sep=3pt,length=3pt] 
      Rectangle[white,sep=3pt,length=3pt]
      Rectangle[white,sep=2pt,length=3pt] Triangle Cap[cap angle=179]},
      shorten > = 0pt,#1]{#2}{#3}{#4}}
  \newcommand{\mbdraw}[5][]{\mess[arrows=->,shorten > = 3pt,shorten < = 3pt, blue,dashed,#1]{#2}{#3}{#4}[#5]}
   
\fi
\ifx\algorithmiccomment\undefined
\else
  \renewcommand{\algorithmiccomment}[1]{{\color{gray}// #1}}
\fi

\IfClassLoadedTF{beamer}{}
{
\ifx\lemma\undefined
  \newtheorem{lemma}[thm]{Lemma}
  \newtheorem{lemma*}[thm]{Lemma}
\else
\fi

\ifx\corollary\undefined
  \newtheorem{corollary}[lemma]{Corollary}
\else
\fi

\ifx\theorem\undefined
  \newtheorem{theorem}[lemma]{Theorem}
  \newtheorem{theorem*}[lemma]{Theorem}
\else
\fi

\ifx\definition\undefined
  \theoremstyle{break}
  \theorembodyfont{ \normalfont}
  \newtheorem{definition}[lemma]{Definition}
  \newtheorem{definition*}[lemma]{Definition}
\else

\fi

\theoremstyle{remark}

\ifx\remark\undefined
  \newtheorem{remark}[lemma]{Remark}
  \newtheorem{remark*}[lemma]{Remark}
\else

  \fi

  \ifx\example\undefined 
  \newtheorem{example}[thm]{Example}
  \newtheorem{example*}[thm]{Example}
\else

\fi

\ifx\notation\undefined
\newenvironment{notation}[1]{
  \vspace{5pt}
  \setlength{\parindent}{-10pt}%
  \setlength{\leftskip}{10pt}%
  \setlength{\rightskip}{0pt}%
    \textbf{#1} :}
{\par\vspace{5pt}}
\fi
\ifx\proof\undefined
  \newenvironment{proof}[1]{
  \noindent
  \textbf{proof: }
 \begin{adjustwidth}{0.25cm}{0.25cm}}
  {\hfill$\Box$
 \end{adjustwidth}}
\fi

\DeclareDocumentEnvironment{mydefinition}{O{} O{ }}{%
  \begin{definition}[{#1}]%
  \ifx#2 
  \else
    \label{def:#2}%
  \fi
}{\end{definition}}
\DeclareDocumentEnvironment{mylemma}{O{ }}
{\restatable{lemma}{lem:#1}%
\ifx#1 
\else
  \label{lem:#1}%
\fi}
{\endrestatable}
\DeclareDocumentEnvironment{mycorollary}{O{ }}
{\restatable{corollary}{cor:#1}%
  \ifx#1 
  \else
    \label{cor:#1}%
  \fi}
  {\endrestatable}
\DeclareDocumentEnvironment{mytheorem}{O{ }}
{\restatable{theorem}{thm:#1}%
\ifx#1 
\else
  \label{thm:#1}%
\fi}
{\endrestatable}
\DeclareDocumentEnvironment{myremark}{O{ }}
{\begin{remark}%
  \ifx#1 
  \else
    \label{rem:#1}%
  \fi}
  {\end{remark}}

\newenvironment{proof-sketch}{\begin{proof}{\hspace{-5pt}\textbf{\color{gray}[sketch]}}}{\end{proof}}
}
\makeatletter
\newcommand{\myrecall}[1]{\csname#1\endcsname*}
\makeatother
\definecolor{todocol}{RGB}{249,226,52}
\newcommand{\toudoux}[1]{\todo[color=todocol,inline]{#1}}

\definecolor{commentColRomain}{RGB}{29, 182, 82}
\definecolor{commentColAnca}{RGB}{182, 29, 82}
\definecolor{commentColGregoire}{RGB}{148, 0 211}
\newcommand{\pcomment}[3]{{\small\color{#1}#3\{#2\}}}
\newcommand{\anca}[1]{\pcomment{commentColAnca}{#1}{A}}

\newcommand{\romain}[1]{{\pcomment{commentColRomain}{#1}{R}}}
\newcommand{\gregoire}[1]{{\pcomment{commentColGregoire}{#1}{G}}}

\newcommand{\set}[1]{\left\{#1\right\}}

\newcommand{\buf}{\mathsf{bf}}
\newcommand{\leqb}{<_\Nn}
\newcommand{\sync}{\mathit{sync}}
\newcommand{\bS}{\bar{S}}
\newcommand{\mss}{\text{\texttt{marked}}}

\newcommand{\astgen}[1][]{\ast_{\scriptscriptstyle #1}}
\newcommand{\astn}{\astgen[\Nn]}
\newcommand{\astm}{\astgen[\mb]}
\newcommand{\astp}{\astgen[\ptp]}
\newcommand{\lehb}{\le_{\text{hb}}}
\newcommand{\lhb}{<_{\text{hb}}}

\newcommand{\skel}{\texttt{skel}}

\DeclareMathOperator*{\bigasync}{\prod}

\usepackage{quotes}
\knowledgeconfigure{diagnose line=true}
\knowledgedirective{standard}{autoref,style=standard,intro style=intro standard}
\knowledgedirective{notion}{autoref,style=standard,intro style=intro standard}
\knowledgeconfigure{quotation}
\knowledgestyle{standard}{color=Blue4}
\knowledgestyle{intro standard}{emphasize} 
\knowledgedefault*{standard}

\knowledge{notion}
 | CFM
 | CFMs
 | communicating finite state machine
 | communicating finite-state machine
 | Communicating Finite State Machine
 | Communicating Finite-State Machine
 | communicating finite-state machines

\knowledge{notion}
 | global states

\knowledge{notion}
 | trace
 | traces

\knowledge{notion}
 | execution
 | executions

\knowledge{notion}
 | Message Sequence Chart
 | message sequence chart
 | MSC
 | message sequence charts
 | msc
 | MSCs

\knowledge{notion}
 | viable
 | $\Nn$-viable
 | $\mb$-viable
 | $\ptp$-viable
 | $\Nn$-viability
 | $\ptp$-viability
 | \Nn-viable
 | \mb-viable
 | \ptp-viable
 | \Nn-viability
 | \ptp-viability

\knowledge{notion}
 | unmatched
 | matched

\knowledge{notion}
 | equivalent
 | equivalence
 | \equiv
 | $\equiv$
 | \myequiv
 | \not\myequiv
 
 \knowledge{notion}
 | communication graph
 | Communication Graph
 | $\Nn$-communication graph
 | $\mb$-communication graph
 | \Nn-communication graph
 | \mb-communication graph

\knowledge{notion}
 | $\Nn$-exchange
 | $\Nn$-exchanges
 | $\mb$-exchanges
 | $\mb$-exchange
 | \Nn-exchange
 | \Nn-exchanges
 | \mb-exchanges
 | \mb-exchange
 | $\ptp$-exchange
 | $\ptp$-exchanges
 | exchanges
 | exchange

 \knowledge{notion}
 | synchronizable
 | $\Nn$-synchronizable
 | $\mb$-synchronizable
 | $\mb$-synchronizability
 | \Nn-synchronizable
 | \mb-synchronizable
 | \ptp-synchronizable
 | $\ptp$-synchronizable
 | \mb-synchronizability

 \knowledge{notion}
 | $\Nn$-synchronous
 | $\mb$-synchronous
 | \Nn-synchronous
 | \mb-synchronous

\knowledge{notion}
 | marked send
 | marked
 | ms
 | marked sends
 | ms-sequence
 | ms-sequences
 | ms sequences
 | marked send sequence
 | marked send sequences

\knowledge{notion}
 | well-labeled
 | well-labeling
 | well-labelings

\knowledge{notion}
 | direct arc
 | direct arcs
 
\knowledge{notion}
 | indirect arc
 | indirect arcs

\knowledge{notion}
| channels
| channel

\knowledge{notion}
 | peer-to-peer semantics
 | semantics
 | mailbox semantics

\knowledge{notion}
 | mailbox
 | mailboxes

\knowledge{notion}
 | buffers
 | buffer

\knowledge{notion}
 | process network
 | process networks

\knowledge{notion}
 | buffer order
 | mailbox order

\knowledge{notion}
 | $\Nn$-valid
 | $\mb$-valid
 | \Nn-valid
 | \mb-valid

\knowledge{notion}
 | $\Nn$-atomic
 | $\ptp$-atomic
 | $\mb$-atomic
 | \Nn-atomic
 | \ptp-atomic
 | \mb-atomic
 | atomic
 | atomicity

\knowledge{notion}
 | mailbox-similarity
 | mailbox-similar

\knowledge{notion}
 | many-to-one

 \knowledge{notion}
 | deaf

\knowledge{notion}
 | process order

\knowledge{notion}
 | $R$-closed

\knowledge{notion}
 | $R$-diamond

 \knowledge{notion}
 | active

\let\myequiv\equiv

\renewcommand{\equiv}{\mathop{\boldsymbol{"\myequiv"}}}

\def\eg{{\em e.g.}}
\def\cf{{\em cf.}}

\renewcommand{\gregoire}[1]{\undefined}
\renewcommand{\anca}[1]{\undefined}
\renewcommand{\romain}[1]{\undefined}
\renewcommand{\toudoux}[1]{\undefined}

\begin{document}

\title[Synchronizable mailbox communication]{An automata-based approach for synchronizable mailbox communication\rsuper*}
\titlecomment{{\lsuper*}A preliminary version of this paper appeared at CONCUR'24.}

\thanks{%
  This work was (partially) supported by the grant ANR-23-CE48-0005 of the French National
  Research Agency ANR (project PaVeDyS)%
}

\author[R.~Delpy]{Romain Delpy\lmcsorcid{0009-0006-0716-3787}}
\author[A.~Muscholl]{Anca Muscholl\lmcsorcid{0000-0002-8214-204X}}
\author[G.~Sutre]{Grégoire Sutre\lmcsorcid{0009-0004-3839-0005}}

\address{LaBRI, Univ. Bordeaux, CNRS, France}

\begin{abstract}
We revisit finite-state communicating systems with round-based communication under mailbox semantics. Mailboxes correspond to one FIFO buffer per process (instead of one buffer per pair of processes in peer-to-peer systems). Round-based communication corresponds to sequences of rounds in which processes can first send messages, then only receive (and receives must be in the same round as their sends).
A system is called synchronizable if every execution can be re-scheduled into an equivalent execution that is a sequence of rounds.
Previous work mostly considered the setting where rounds have fixed size.
Our main contribution  shows
that the problem whether a mailbox communication system complies with
the round-based policy, with no size limitation on rounds, is
\PSPACE-complete.
For this we use a novel automata-based approach, that also allows to
determine the precise complexity (\PSPACE)
of several questions considered in previous literature.
\end{abstract}

\maketitle

\section{Introduction}
Message-passing is a key synchronization feature for concurrent programming and
distributed systems.
In this model,  processes running asynchronously synchronize by exchanging
messages over unbounded channels.
The usual semantics is based on peer-to-peer communication, which
is very popular for reasoning about telecommunication
protocols.
More recently, mailbox communication received increased attention
because of its usage in multi-thread programming, as provided  by 
languages like Rust or Erlang.
Mailbox communication means that every process has a single incoming
communication buffer on which incoming messages from other processes are
multiplexed  (a mailbox).

Message-passing programs are well-known to be challenging for 
formal verification since they can easily simulate Turing machines with
unbounded channels. 
Some approximation techniques can help to recover decidability.
Among the best known  approaches are 
lossy channel systems~\cite{AbdullaJ96,FinkelS01} and
partial-order methods~\cite{KuskeM21}.
The latter tightly relate to (high-level) message sequence charts (HMSC),
a communication formalism capturing multi-party session types~\cite{Stutz23,LiSWZ23,MMSZ21}.
An HMSC  protocol is  a graph with nodes labelled by communication
scenarios, a.k.a.~message sequence charts.
Processes still evolve asynchronously, so that the division into
nodes cannot be enforced by global synchronization.
Such round-based communication is actually  quite frequent in distributed
computing, for example as building block in the Heard-Of model~\cite{Charron-BostS09}.
Often a distributed protocol consists of several rounds, where each
round first has a phase where processes only send messages, then a phase
where they only receive.
We refer to such rounds as $\texttt{sr}$-rounds.

Recently $\texttt{sr}$-round-based communication and mailbox communication were
considered together  in~\cite{bouajjani2018completeness}.
It turned out  that this combination is
very attractive for formal verification. 
The  paper~\cite{bouajjani2018completeness} proposed a model where $\texttt{sr}$-rounds have fixed size, and
showed that control-state reachability in this model becomes decidable  (in \PSPACE).
The question whether a system complies with the $\texttt{sr}$-round
model with given round size was shown to be decidable in~\cite{digiusto20fossacs}.
It is also known how to decide if a system
complies with the $\texttt{sr}$-round
model when the round size is not known in advance~\cite{GiustoLL23}.
All these properties motivate a  genuine interest in the $\texttt{sr}$-round model
on top of mailbox communication.
A bit surprisingly, apart from control-state reachability, similar questions were shown
to be undecidable for peer-to-peer communication~\cite{genest2007communicating}.

In this paper we revisit the framework of
\cite{bouajjani2018completeness} and propose an automata-based
approach to deal with systems complying with the $\texttt{sr}$-round
mailbox model (we refer to this property as
\emph{$\mb$-synchronizability}). 
Importantly, we do not impose any size restriction on the rounds, as
in previous works.
This makes sense, because even when we can infer an upper bound on the size as
in~\cite{GiustoLL23}, this upper bound is exponential in the number of
processes, so its  practical use is somewhat limited.
We  establish that the complexity of all problems listed below is
\PSPACE-complete for \emph{$\mb$-synchronizable} systems:
\begin{itemize}
\item Global-state reachability (Theorem~\ref{thm:reach-for-mb-sync}).
 \item Model-checking against a reasonable class of regular properties
   (Theorem~\ref{thm:Rclosed-model-checking}).
   \item  Check if a peer-to-peer system can be simulated as a mailbox
     system (modulo rescheduling executions, Theorem~\ref{thm:mailbox-similarity-pspace}).
 \end{itemize}
 Our main result is that $\mb$-synchronizability can be checked in \PSPACE\
     (Theorem~\ref{thm:mb-synchronizability}), the complexity being tight.
An interesting byproduct of our results is that when we fix the number
of processes all the problems above can be solved in \PTIME\ (actually
\NLOGSPACE).

\paragraph{Comparison with related work.}
Our technique helps to establish
the precise complexity of several problems considered in the papers
mentioned above.
To be precise, our definition of $\texttt{sr}$-round mailbox model
(\emph{$\mb$-synchronizability}) slightly differs from the one used in
\cite{bouajjani2018completeness,digiusto20fossacs,GiustoLL23} (but
coincides with a variant introduced in \cite{BolligGFLLS21Concur}).
The latter paper uses a partial-order variant of PDL (LCPDL) to show an
\EXPTIME\ upper bound for the synchronizability problem for their
notion of synchronizability.
Using MSO logic and special tree-width, the paper
\cite{BolligGFLLS21Concur} also shows that checking if a 
system is synchronizable with fixed round size is decidable.
Knowing if a round size exists is shown to be decidable with
elementary complexity in
\cite{GiustoLL23}, without exact  bounds.

\paragraph{Outline.}
Section~\ref{sec:def} introduces the model and the questions considered in this
paper. Section~\ref{sec:reach}
discusses the reachability problem for synchronizable systems, using an automata based technique.
Then we show how our techniques allow to check some regular properties in
Section~\ref{sec:properties}. In
Section~\ref{sec:ptp_mb}, we consider the question if the behaviors of a system are
the same in the mailbox and in the peer-to-peer semantics.
Section~\ref{sec:sync} presents the main contribution, showing that the question
if a system is \mb-synchronizable is \PSPACE-complete.
Finally, in Section~\ref{sec:comp} we discuss other existing notions of synchronizability and how they
compare to the notion studied in this paper.

\medskip

For convenience, technical terms and notations in the
electronic version of this manuscript are hyper-linked to their 
definitions (\cf~\url{https://ctan.org/pkg/knowledge}).
A preliminary version of this paper was presented at CONCUR 2024~\cite{DBLP:conf/concur/DelpyMS24}.

\section{Message-passing systems and synchronizability}
\label{sec:def}
Throughout the paper,
$\Proc$ denotes a finite non-empty set of \emph{processes}, and
$\Msg$ denotes a finite non-empty set of \emph{message contents}.
We consider here
peer-to-peer communication between distinct processes.
Formally,
the set of (communication) ""channels"" is the set $\Ch$
of all pairs $(p, q) \in \Proc \times \Proc$ such that $p \neq q$, and
the set of (communication) \emph{actions} is 
  $\Act = \{p!q(m), q?p(m) \mid (p, q) \in \Ch, m \in \Msg\}$. 
An action $p!q(m)$ denotes a \emph{send} by $p$ of message $m$ to $q$ and
an action $p?q(m)$ denotes a \emph{receive} by $p$ of message $m$ from $q$.
In both cases, the process performing the action is $p$.
Throughout the paper,
we let $S$ and $R$ denote the sets of send actions and receive actions,
formally,
$S = \set{p!q(m) \mid (p,q) \in \Ch, m \in \Msg}$ and
$R = \set{p?q(m) \mid (q,p) \in \Ch, m \in \Msg}$.

\smallskip

A communicating finite state machine \cite{brand1983communicating} is
a finite set of processes that exchange messages,
each process being given as a finite LTS.
Recall that a (finite) \emph{labeled transition system},
\emph{LTS} for short,
is a quadruple $(L, A, \to, i)$ where
$L$ is a (finite) set of \emph{states},
$A$ is a finite alphabet,
${\to} \subseteq L \times A \times L$ is a set of \emph{transitions}, and
$i \in L$ is an \emph{initial} state.
We will sometimes consider LTS without initial state.
In the following definition,
$\Act_p$ denotes the set of actions $a \in \Act$ performed by $p$.

\begin{mydefinition}[Communicating Finite-State Machine]
  \AP
  A ""CFM"" is a tuple $\Aa = (\Aa_p)_{p\in \Proc}$,
  where each $\Aa_p$ is a finite LTS
  $\Aa_p = (L_p, \Act_p, \to_p, i_p)$.
  States in $L_p$ are called \emph{local states}.
  The \emph{size} of $\Aa$ is defined as $\sum_{p \in\Proc} (|L_p|+|\to_p|)$.
\end{mydefinition}

In this paper,
we mainly study and compare two semantics of communication:
peer-to-peer and mailbox.
These two semantics differ in the implementation of the communication network.
\AP In the ""peer-to-peer semantics"",
each channel $(p, q)$ is implemented by a dedicated fifo buffer.
This is the classical semantics for "communicating finite-state
machines"~\cite{brand1983communicating}. 
In the \emph{mailbox semantics},
each process $q$ is equipped with a fifo buffer that acts as a ""mailbox"":
all messages towards $q$ are enqueued in this buffer.
Put differently,
the channels $(p, q)$ with same receiver $q$ are multiplexed into a single buffer.

We define both semantics of "CFM" jointly, by viewing  "channels"
and "mailboxes" as (fifo) message "buffers":

\begin{mydefinition}[Process network]
  \AP
  A ""process network"" over $\Proc$
  is a pair $\Nn = (\Bufset, \buf)$ where $\Bufset$ is a finite set of fifo ""buffers""
  and $\buf : \Ch \rightarrow \Bufset$ is a map that assigns a
  "buffer" to each "channel".
\end{mydefinition}

The "peer-to-peer semantics" is induced by the "process network"
$\ptp = (\Bufset, \buf)$ where $\Bufset = \Ch$ and $\buf$ is the identity.
Here, $\Bufset$ coincides with the set of communication "channels".
The "mailbox semantics" is induced by the "process network"
$\mb = (\Bufset, \buf)$ where $\Bufset = \Proc$ and $\buf(p, q) = q$.
Here, $\Bufset$ is a set of "mailboxes", one per process.

\begin{myremark}
For both "peer-to-peer semantics" and "mailbox semantics" 
we have that the "buffer" determines the recipient: $\buf(p, q) = \buf(p', q')$ implies $q = q'$.
 We call such "process networks" ""many-to-one"".
\end{myremark}

Given a "CFM" and a "process network" we define the associated
global transition system:

\begin{mydefinition}[Global transition system]
\AP  Let $\Aa = (\Aa_p)_{p\in \Proc}$ be a "CFM", and
  $\Nn = (\Bufset,\buf)$ be a "process network" over $\Proc$.
  The \emph{global transition system} associated with $\Aa,\Nn$ is
  the LTS
  $\Tt_\Nn(\Aa) = (C_{\Aa},\Act,\rightarrow_{\Aa},c_{in})$ with set of
  configurations $C_\Aa = \Glob \times ((\Ch \times
  \Msg)^*)^{\Bufset}$  consisting
    of ""global states"" $\Glob = \prod_{p\in\Proc}L_p$ (i.e., products of  local states) 
    and "buffer" contents, with $((\ell_p)_{p\in\Proc},(w_{\bufelt})_{\bufelt \in \Bufset}) \xrightarrow{a}_\Aa
      ((\ell_p')_{p\in\Proc},(w'_\bufelt)_{\bufelt \in \Bufset})$ if 
    \begin{itemize}
    \item $\ell_p\xrightarrow{a}_p \ell_p'$ and $\ell_q=\ell_q'$ for
      $q\not=p$, where $p$ is the process performing $a$.
    \item Send actions: if $a = p!q(m)$ then $w'_{\bufelt} = w_{\bufelt} \, ((p, q), m)$
      and $w'_{\bufelt'} = w_{\bufelt'}$ for $\bufelt'\not=\bufelt$, where $\bufelt = \buf(p, q)$.
    \item Receive actions: if $a = p?q(m)$ then $((p, q), m) \, w'_{\bufelt} = w_{\bufelt}$
      and $w'_{\bufelt'} = w_{\bufelt'}$ for $\bufelt'\not=\bufelt$, where $\bufelt = \buf(p, q)$.
    \end{itemize}
    The initial configuration is $c_{in}=((i_p)_{p \in \Proc},\varepsilon^{\Bufset})$.
\end{mydefinition}

\AP An ""execution"" of $\Tt_\Nn(\Aa)$ is a sequence
$\rho=c_0 \act{a_1} c_1 \cdots \act{a_n} c_n$ with $c_i\in C_{\Aa}$
such that $c_{i-1} \xrightarrow{a_i}_\Aa c_{i}$ 
for every $i$.
The sequence $a_1 \cdots a_n$ is the \emph{label} of the "execution".
The "execution" is \emph{initial} if $c_0=c_{in}$.

\begin{myremark}
  Note that in the definition above we added the channel name to the
  message content inserted in a buffer. 
This is to exclude "executions" like $p!q(m) \; q?r (m)$ with $p \not= r$.
Without this addition such "executions" would be allowed in the mailbox
semantics, which is clearly not intended.
\end{myremark}

\begin{mydefinition}[Trace]
  \AP
  A ""trace""  of a "CFM" $\Aa$ over a "process network" $\Nn$ is a sequence $u \in \Act^*$
  such that there exists an initial "execution" of $\Tt_\Nn(\Aa)$
  labelled by $u$.
  The set of all "traces" of $\Aa$ is denoted by $\Ex_\Nn(\Aa)$.
\end{mydefinition}

As we will also need to consider infixes of "executions", we introduce
action sequences which are coherent
w.r.t.~the fifo behavior that we expect from a "process
    network":

\begin{mydefinition}[Viable sequence]
    \AP Let $\Nn=(\Bufset,\buf)$ be a  "process
    network".
    A sequence of actions $v\in\Act^*$  is called ""$\Nn$-viable"" if for every
    "buffer" $\bufelt \in \Bufset$:
    \begin{itemize}
        \item  for every prefix $u$ of $v$, the number of receives from $\bufelt$ in $u$ 
        is less or equal the number of sends to $\bufelt$ in $u$;
        \item for every $k$, if the $k$-th receive from $\bufelt$ in $v$  has label
        $q?p(m)$ then the $k$-th send to $\bufelt$ in $v$ has label $p!q(m)$.
    \end{itemize}
\end{mydefinition}

\noindent 
There is a strong connection between "traces" and "viable" sequences.
For every sequence $u \in \Act^*$,
$u$ is a "trace" of $\Aa$ over $\Nn$ iff
$u$ is "$\Nn$-viable" and $u$ is recognized by $\bigasync_{p \in \Proc} \Aa_p$.
Here,
$\bigasync_{p \in \Proc} \Aa_p$ denotes the asynchronous product of
the LTS $\Aa_p$,
viewed as automata with every state final.

\begin{myremark}[trace-vs-viable]
    It is easy to see that if a sequence is "$\mb$-viable" then it is
    also "$\ptp$-viable".
    In fact,
    for every "process network" $\Nn$,
    we have that "$\Nn$-viability" implies "$\ptp$-viability".
    However, the converse is not true.
    For example, $p_1!p_2(a)\, p_3!p_2(b)\, p_2?p_3(b)$ is "$\ptp$-viable", but not
    "$\mb$-viable" because $b$ is enqueued after $a$ in $p_2$'s
    mailbox, so it cannot be received first.
\end{myremark}

The classical \emph{happens-before} relation~\cite{Lamport78},
frequently used in
reasoning about
distributed systems, orders the actions
of each process and  every ("matched") send action before its matching
receive.
The happens-before relation  naturally associates a partial order with every
"trace", known as \emph{message sequence chart}:

\begin{mydefinition}[Message Sequence Chart]
    \AP An ""MSC""
    over $\Proc$  is an
    $\Act$-labeled partially ordered set $\Msc=(E,\lehb,\lambda)$
    of events $E$, with $\l:E \to \Act$ and $\lehb \, \mathrel{=} (\le_\Proc \cup \msg)^*$ the least partial
    order containing the relations $\le_\Proc$ and $\msg$, which are defined as:
    \begin{enumerate}
        \item For every process $p$, the set
          of events on $p$ is
          totally ordered by $\le_\Proc$, and $\le_\Proc$ is the union of
          these total orders.
        \item \AP$\msg$ is the set of matching send/receive event pairs.
          In particular, $(e,f) \in \msg$ 
          implies  $\l(e)=p!q(m)$ and
            $\l(f)=q?p(m)$  for some $p,q\in \Proc$ and $m\in \Msg$. Moreover,
              $\msg$ is a partial bijection between sends and receives
              such that every receive is paired with a (unique) send.
              A send is called ""matched"" if it is in the
              domain of $\msg$, and \emph{unmatched} otherwise.
     \end{enumerate}
\end{mydefinition}

\begin{figure}
    \centering
    \scalebox{0.7}{\input{figures/msc_example.tex}}\\
    {\footnotesize$p_1!p_2(a)\,p_2!p_3(b)\,p_3?p_2(b)
    \,p_3!p_2(c)\,p_2?p_1(a)$}
    \caption{A sequence and its "MSC".
    {An "unmatched" send action is marked by a special arrowhead, as for message
    $c$.}}
  \label{fig:msc_ex}
\end{figure}

\noindent 
The fifo behavior of message "buffers" implies that not every "MSC" arises
as possible behavior.
We formalise this for any "process network"
$\Nn=(\Bufset,\buf)$ by defining a 
""buffer order""\footnote{This  definition of $\leqb$ is tailored
  for "many-to-one" "process networks", but for simplicity we have chosen
  not to mention the restriction in the definition.
  Note that $\leqb$ is a strict partial order.
}  $\leqb$ on sends to the same "buffer".
Let $e \leqb e'$ if $e,e'$ are of type $p!q$ and $s!r$,
resp., with $\buf(p,q)=\buf(s,r)$, and
\begin{itemize}
\item either $e$ is "matched" and $e'$ is "unmatched",
\item or $(e,f),(e',f') \in\msg$ and $f<_\Proc f'$.
\end{itemize}

\begin{mydefinition}[Valid "MSC"]
Given a "process network" $\Nn$, an "MSC" $\Msc=(E,\lehb,\lambda)$ is called
""$\Nn$-valid"" if the relation $(\lhb \cup \leqb)$
is acyclic. 
\end{mydefinition}

It is easy to see that an "MSC" is $\ptp$-valid
iff "matched" messages on any channel $(p,q)$ never overtake and "unmatched"
sends by $p$ to $q$ are $\le_\Proc$-ordered after the "matched" sends.
An "MSC" is "$\mb$-valid" iff for any sends $s \lhb s'$ to
the same process, either they are both "matched" and their receives
satisfy $r<_\Proc r'$, or $s'$ is "unmatched".
Figure~\ref{fig:msc_ex} shows an  "$\mb$-valid" "MSC".
An "$\mb$-valid" "MSC" is the same as an "MSC" obtained from
a "trace" that satisfies \emph{causal
  delivery} in \cite{bouajjani2018completeness}, and 
it is called
\emph{mailbox MSC} in~\cite{BolligGFLLS21Concur}.

\bigskip

If $u=u[1] \cdots u[n]$ is a "$\ptp$-viable" sequence of actions then we
can associate an "MSC" with $u$ by setting $\mscn(u)=(E,\lehb,\l)$ with
$E=\{e_1,\dots,e_n\}$, $\l(e_i)=u[i]$, and the orders defined as
expected:
\begin{itemize}
    \item $e_i \leq_\Proc e_j$ if $u[i]$ and $u[j]$ are performed by the same process and  $i \le j$.
    \item $(e_i,e_j) \in \msg$ if there exists $k \geq 1$ and a "buffer" $\bufelt \in\Ch$
      such that $u[i]$ is the $k$-th send to $\bufelt$ and $u[j]$ is the $k$-th
      receive from $\bufelt$.
    \end{itemize}
Note that $\mscn(u)$ only depends (up to isomorphism) on the projection of $u$ on each process.

\noindent
 \emph{Caveat:} Throughout the paper we switch between reasoning on "$\Nn$-viable"
      sequences (when we use automata) and their associated "MSC"
      (when we use partial orders).
      So when we refer to a position in a ("viable") sequence $u$  we often see it
      directly as an event of  $\mscn(u)$, without further
      mentioning it.
      
    \begin{myremark}[viable-vs-valid]
      By definition, for any "$\Nn$-viable" sequence $u$ the associated
     "MSC"  $\mscn(u)$ is "$\Nn$-valid".
     For the converse, if the "process network" is "many-to-one" and
     the "MSC" $\Msc$ is "$\Nn$-valid" then every (labelled)
     linearization of the partial order $(\lhb \cup \leqb)^*$ of
     $\Msc$ %
     is "$\Nn$-viable".
     Indeed,
     all receives from the same buffer are totally ordered by $\le_\Proc$
     when the "process network" is "many-to-one",
     and the corresponding sends are ordered in the same way because of the "buffer order".
For example, the sequence shown in Figure~\ref{fig:msc_ex} is
"$\mb$-viable", but  $p_2!p_3(b)\, p_3?p_2(b)\, p_3!p_2(c)\,
p_1!p_2(a)\, p_2?p_1(a)$ is not.
\end{myremark}

    For a "process network" $\Nn$ and a "CFM" $\Aa$ we write
$\mscn_\Nn(\Aa) =\{\mscn(u) \mid u \in \Ex_\Nn(\Aa)\}$  for the set
of "MSCs" associated with initial "executions" of $\Aa$.
By Remark~\ref{rem:viable-vs-valid},
the set
$\mscn_\Nn(\Aa)$ consists only of "$\Nn$-valid" "MSCs".
The next definition introduces an equivalence relation $\equiv$ on
"CFM" "traces" that is ubiquitous in this
paper.
Two "traces" are equivalent up to commuting adjacent
actions that are neither performed by the same process, nor a
matching send/receive pair: 

\begin{mydefinition}[Equivalence "$\equiv$"]
   \AP Two "$\ptp$-viable" sequences $u,v \in \Act^*$ are called
""equivalent"" if
$\mscn(u)=\mscn(v)$ (up to isomorphism), and we write $u \equiv v$ in this case.
  \end{mydefinition}

\begin{myremark}[equiv-trace]
  Two "$\ptp$-viable" sequences are "equivalent" iff
  they have the same projection on each process.
\end{myremark}

\begin{myremark}[non-viable]
  If $u,v \in \Act^*$ are both "$\Nn$-viable" with $u \equiv v$, then $u
  \in \Ex_\Nn(\Aa)$ iff $v \in
  \Ex_\Nn(\Aa)$.
  However, $\equiv$ does not preserve "$\Nn$-viability".
  Take as example $p!q(a)\,r!q(b)\,q?p(a) \equiv$  $r!q(b)\,p!q(a)\,q?p(a)$, and
  observe that 
the left-hand side is "$\mb$-viable", while the right-hand side is not.
Another example is the sequence $u=p_1!p_2(a)\,p_2!p_3(b)\,p_3?p_2(b)
  \, p_3!p_2(c)\,p_2?p_1(a) $  in Figure~\ref{fig:msc_ex}: it is
  "$\mb$-viable", but $p_3!p_2(c)\,p_1!p_2(a)\,p_2!p_3(b)\,p_3?p_2(b)
  \,p_2?p_1(a)$ which is equivalent to $u$ is not.

\end{myremark}

For the rest of the section $\Nn=(\Bufset,\buf)$ always refers to a
"process network". 
In order to be able to cope with partial "executions"  we start by
observing that  "unmatched" sends to a "buffer" restrict 
the product of "$\Nn$-viable" sequences.
Let $u$ and $v$
    be two "$\Nn$-viable" sequences. The product $u \astn v$
    is defined if for every "buffer" $\bufelt\in\Bufset$, if there is an "unmatched" send to $\bufelt$ in $u$,
    then there is no receive from $\bufelt$ in $v$.
    When it is defined, $u \astn v$ is equal to $uv$. 
  Note that the partial binary operation $\astn$ is associative. %
  Moreover, if $u_1 \astn \cdots u_i \astn \cdots u_j \astn u_{j+1} \cdots u_n$ is
  defined then $u_1 \astn \cdots u_i \astn u_{j+1} \cdots u_n$ is also
  defined, for every $i<j$.
  Note also that, when it is defined, the $\astn$-product of two "$\Nn$-viable" sequences is
  "$\Nn$-viable".

\begin{mydefinition}[Exchanges, synchronizability]
  \begin{enumerate} 
    \item An ""$\Nn$-exchange"" is any "$\Nn$-viable" sequence $w \in S^* R^*$
    (i.e., sends followed by receives).
    \item An "$\Nn$-viable" sequence  $u$ is called ""$\Nn$-synchronous"" if it is a
      $\astn$-product of "$\Nn$-exchanges".
      It is called ""$\Nn$-synchronizable""\footnote{A different notion of \emph{synchronizability} appears in~\cite{bouajjani2018completeness,digiusto20fossacs}.
      We discuss the difference between the two notions in Section~\ref{sec:comp}.} if $u \equiv v$ for some 
"$\Nn$-synchronous" sequence $v$.
\item A "CFM" $\Aa$ is "$\Nn$-synchronizable" if all its "traces" $u \in
\Ex_\Nn(\Aa)$ are "$\Nn$-synchronizable".
  \end{enumerate}
\end{mydefinition}

\noindent 
We end this section with a comparison between synchronizability 
for "peer-to-peer semantics" and "mailbox semantics".
These two notions are incomparable, in general.
First, "$\mb$-synchronizability" does not imply $\ptp$-synchronizability simply because
a system under $\mb$-semantics has less "executions" than under
$\ptp$-semantics.
Conversely, the "execution" 
\[p_1!p_3(a) \, p_2!p_1(b) \,
p_1?p_2(b) \, p_1!p_2(c) \, p_2?p_1(c) \, p_3!p_2(d) \, p_3?p_1(a)
\]
depicted in Figure~\ref{fig:msc_ex2}
is "$\mb$-viable" and "$\ptp$-synchronizable", but not
"$\mb$-synchronizable".
\begin{figure}
    \centering
    \scalebox{0.7}{\input{figures/msc_example2.tex}}
    \caption{%
      "MSC" of an "$\mb$-viable" sequence that is "$\ptp$-synchronizable",
      but not "$\mb$-synchronizable".
    }
  \label{fig:msc_ex2}
\end{figure}
The above execution is "$\ptp$-synchronizable" because the "unmatched" send
$p_3!p_2(d)$ can be executed as first event in the $\ptp$-semantics.
However, as we will see in Section~\ref{sec:atomic}, this is not the case in the
$\mb$-semantics.

Finally, we note that $\ptp$-synchronizability  was shown to be undecidable in
\cite{BolligGFLLS21Concur}.

\section{Reachability for $\mb$-synchronizable systems}
\label{sec:reach}
We start this section by showing that
state reachability for "$\mb$-synchronizable" "CFMs" is
\PSPACE-complete.
The decidability (in exponential time) for "$\mb$-synchronizable" "CFMs"
can already be inferred from~\cite{BolligGFLLS21Concur} using the partial order
logic LCPDL.
The main point of this section is to introduce an
automata-based approach to deal with "$\mb$-synchronizable" "CFMs". 
Although the  set of "$\mb$-synchronous" "traces" of a "CFM" is not regular in general,
the projection of this set on ("marked") send actions turns out to be
regular.
This crucial property is used later as a basic ingredient by our
algorithm for deciding "$\mb$-synchronizability".

We start with an important observation saying that
"$\mb$-synchronizability" allows to focus
on send actions.
However, "unmatched" and "matched" sends need to be distinguished.
So we introduce an extended alphabet $\bS=\set{\bar{s} \mid s \in S}$.
Sequences over $S \cup \bar{S}$ will be referred to as ""marked send sequences"" ("ms-sequences" for short).
For any "$\mb$-viable" sequence $u$, we annotate every "unmatched" send
$p!q(m)$ in $u$ by $\overline{p!q(m)}$ and we denote by $\mss(u)$ the
sequence obtained in this way.
For example, for $u = p!q(m) p!r(m') r?p(m')$ we have
$\mss(u)=\overline{p!q(m)} p!r(m') r?p(m')$.
\AP The  "ms-sequence" $\ms(u)$ associated
  with an "$\mb$-viable" sequence $u$ is the
  projection of $\mss(u)$ on $S \cup \bS$.

\begin{mylemma}[marked]
  \begin{enumerate}
  \item  For any  "$\mb$-exchanges" $u,u'$ with $\ms(u) = \ms(u')$,
    we have $u \equiv u'$.
    \item For any "$\mb$-exchange" $u=v v'$ with $v \in S^*, v' \in
      R^*$,  we define 
      $\hat{u}=v v''$ with $v''$ obtained from $v'$ by ordering the
      receives %
      as their matching sends in $v$.
      Then $\hat{u}$ is "$\mb$-viable" and $u \equiv \hat{u}$.
  \end{enumerate}
\end{mylemma}

\begin{proof}
For item 1, as $\ms(u) = \ms(u')$ and $u,u'$ are both "$\mb$-viable", we get that
   for each process $p$, the sequence of receives by $p$ in $u$ and
   $u'$, resp., are the same.
   We derive from $u, u' \in S^* R^*$ that
   $u$ and $u'$ have the same projection on each process,
   and thus $u \equiv u'$. For item 2 it is easy to check that $\hat{u}$ is "$\mb$-viable", hence $u
   \equiv \hat{u}$ by item 1.
 \end{proof}

\begin{myremark}
  It is worth noting that Lemma~\ref{lem:marked} does not hold anymore
  under $\ptp$-semantics.
  For example, the two "$\ptp$-exchanges" $u=p_1!p_2(a)\, p_3!p_2(b)\, p_2?p_3(b)\,
  p_2 ? p_1 (a)$ and $\hat{u}=p_1!p_2(a)\, p_3!p_2(b)\, p_2?p_1(a)\,
  p_2 ? p_3 (b)$  have the same "marked" sequence, but they are not
  equivalent.
  This is the main reason why our decidability results don't carry
  over to the $\ptp$-semantics.
 \end{myremark}

\subsection*{Executable "$\mb$-exchanges"}
We now show how to check if an "ms-sequence" corresponds to an 
executable "$\mb$-exchange" of a "CFM" $\Aa$.
Since we use the same construction also for the model-checking problem in
Section~\ref{sec:properties} we give a more general
formulation below.

Given an "$\mb$-viable" sequence $u$ and two sets $D, D' \subseteq \Proc$,
we write $D \executable{u} D'$ if
no process from $D$ receives any message in $u$, and
$D'$ contains $D$ and those processes $q$ such that $u$
has some "unmatched" send to $q$.
We refer to processes in $D, D'$ as ""deaf"" processes.
It is routinely checked that,
for every "$\mb$-viable" sequences $u_1, \ldots, u_n$,
the product $u_1 \astm \dots \astm u_n$ is defined
iff
$D_0 \executable{u_1} D_1 \cdots \executable{u_n} D_n$
for some sets $D_0, \ldots, D_n$.

\begin{mydefinition}[$R$-diamond]
  Let $\Aa=(L, S \cup \bS \cup R, \act{}_\Aa)$ be an LTS.
We say that $\Aa$ is ""$R$-diamond"" 
  if for all states $\ell,\ell' \in L$
and all \emph{receives} $a,a' \in R$ performed by different processes,
we have $\ell \act{aa'}_\Aa \ell'$ iff $\ell \act{a'a}_\Aa \ell'$.
\end{mydefinition}

For any states $\ell,\ell'$ of $\Aa$, sets $D,D' \subseteq \Proc$ and
"$\mb$-viable" sequence $u$, we write
$
(\ell,D) \executable{u}_\Aa (\ell',D')
$
if $\ell \act{\mss(u)}_\Aa \ell'$ and $D \executable{u} D'$.
  The next lemma shows how to adapt an "$R$-diamond" LTS to
  work on "ms-sequences" instead of "$\mb$-synchronous" sequences (a similar
  idea appears in \cite{GiustoLL23}):

\begin{mylemma}[automaton-ms]
  Assume that $\Aa=(L, S \cup \bS \cup R, \act{}_\Aa)$ is an "$R$-diamond" LTS.
  Then we can construct an LTS with $\varepsilon$-transitions
  $\Aa_\sync=((L \cup L^3) \times 2^\Proc, S \cup \bS, \act{}_\sync)$
  such that for any $v \in (S \cup \bS)^*$,
  states $\ell,\ell' \in L$, and sets $D,D' \subseteq \Proc$:
  \[
    (\ell,D) \act{v}_\sync (\ell',D')
    \quad \text{ iff } \quad
    \exists u \text{ "$\mb$-synchronous" } \text{ s.t. }
    v=\ms(u) \text{ and } (\ell,D) \executable{u}_\Aa (\ell',D')
  \]
\end{mylemma}

\begin{proof}
  The LTS $\Aa_\sync$ has the following transitions, for any $\ell,\ell' \in
  L$, $D,D' \subseteq \Proc$, $a \in S \cup \bS$:
  \[
  \begin{cases}
    (\ell,D) \act{\varepsilon}_\sync (\ell,\hat{\ell},\hat{\ell},D) & \text{ for any } \hat{\ell} \in L\\
    (\ell,\ell',\hat{\ell},D) \act{a}_\sync (\ell_1,\ell'_1,\hat{\ell},D) & \text{ if } a =p!q(m), q \notin
    D, \ell \act{a}_\Aa \ell_1,  \ell'\act{q?p(m)}_\Aa \ell'_1 \\
    (\ell,\ell',\hat{\ell},D) \act{a}_\sync (\ell_1,\ell',\hat{\ell}, D') & \text{ if } a=
    \overline{p!q(m)}, \ell \act{a}_\Aa \ell_1,     D'=D \cup \{q\}\\
    (\ell,\ell',\hat{\ell}, D) \act{\varepsilon}_\sync (\ell',D) & \text{ if } \ell=\hat{\ell}
  \end{cases}
 \]

\noindent 
In other words, from a state $(\ell,D) \in L \times 2^{\Proc}$ the LTS
$\Aa_\sync$ first guesses a ``middle'' state $\hat{\ell}
\in L$ for the current "exchange", as the state reached after the
sends.
Then it switches to state $(\ell,\hat{\ell},\hat{\ell}, D)$.
The first component and the second component track sends and their
matching receives (if "matched") in a ``synchronous'' fashion.
The LTS $\Aa_\sync$ also guesses the end of the current "$\mb$-exchange", checking that
the first component has reached the middle state $\hat{\ell}$ guessed originally.
The claimed property of $\Aa_\sync$ follows from
Lemma~\ref{lem:marked} (2) and from  $\Aa$ being "$R$-diamond".
\end{proof}

Fix now a "CFM" $\Aa$.
We abusively use the same notation $\executable{}_{\Aa}$ as above for LTS:
for any "global states" $g, g' \in G$ of $\Aa$,
sets $D,D'\subseteq \Proc$ and
"$\mb$-viable" sequence $u$,
we write $(g,D) \executable{u}_{\Aa} (g',D')$ if
$u$ labels an "execution" in $\Tt_\mb(\Aa)$ from the
configuration $(g,\varepsilon^\Bufset)$ to some configuration
$(g',(w_{\bufelt})_{\bufelt \in \Bufset})$,
and $D \executable{u} D'$.
We obtain from the previous lemma that:

\begin{mylemma}[aut-for-exchange]
  Let $\Aa$ be a "CFM",
   $g, g' \in G$  two "global states" of $\Aa$,
  and $D,D'\subseteq \Proc$ two sets of processes. 
  One can construct automata $\Bb,\Cc$  with $O(|G|^3\times
  2^{|\Proc|})$ states such that 
  \begin{eqnarray*}
    L(\Bb) &=&\set{v \in (S \cup \bar{S})^* \mid \exists u \text{
    "$\mb$-exchange" } \text{ s.t. } v=\ms(u) \text{ and } (g,D)
               \executable{u}_{\Aa} (g',D')}\,,\\
    L(\Cc) &=& \set{v \in (S \cup \bar{S})^* \mid \exists u \text{
    "$\mb$-synchronous" } \text{ s.t. } v=\ms(u) \text{ and } (g,D)
               \executable{u}_{\Aa} (g',D')}\,.%
  \end{eqnarray*}
\end{mylemma}
\begin{proof}
  Assume that $\Aa = (\Aa_p)_{p \in \Proc}$.
  Let $\Qq$ denote the asynchronous product $\bigasync_{p \in \Proc} \bar{\Aa}_p$,
  where each $\bar{\Aa}_p$ is the LTS obtained from $\Aa_p$ by
  adding a transition $\ell_p\xrightarrow{\bar{s}}_p \ell_p'$ for each transition $\ell_p\xrightarrow{s}_p \ell_p'$ with $s \in S$.
  Note that $\Qq$ is "$R$-diamond".
  Moreover,
  it is routinely checked that,
  for every "$\mb$-viable" sequence $u$,
  the relation $\executable{u}_{\Aa}$ coincides with the relation $\executable{u}_{\Qq}$.

  For $\Cc$ we take the
  automaton $\Qq_\sync$ constructed according to
  Lemma~\ref{lem:automaton-ms},
  and set the initial state to $(g,D)$ and the final state to $(g',D')$.
  For $\Bb$, we need to tinker a bit with $\Qq_\sync$ to ensure that we read only one "exchange".
  So we remove all transitions from/to states in $L \times 2^\Proc$ except
  the transitions from $(g,D)$, which we set as initial, and
  the transitions to $(g',D')$, which we set as final.
  If $(g,D)=(g',D')$ then we make two different states for the initial and the final one.
\end{proof}
Using Lemma~\ref{lem:aut-for-exchange} we   establish the
upper bound of the global-state reachability problem  for  "$\mb$-synchronizable"
"CFMs" (the lower bound is straightforward).
By global-state reachability
we mean the existence of a reachable configuration with a specified
global state.

\begin{mytheorem}[reach-for-mb-sync]
  The global-state reachability problem for "$\mb$-synchronizable" "CFMs"
  is \PSPACE-complete.
\end{mytheorem}

\begin{proof}
  Note first that if $\Aa$ is a "CFM" and $u,v$ two "$\mb$-viable" sequences $u,v$
  with $u \equiv v$ then $c_{in} \act{u}_\Aa c$ implies that  $c_{in}
  \act{v}_\Aa c'$ for some $c'$ with the same global state as $c$.
  Since we assume that the "CFM" is "$\mb$-synchronizable" we can
  choose $v$ to be   "$\mb$-synchronous".
  Thus we can use  automaton $\Cc$ from
  Lemma~\ref{lem:aut-for-exchange} to show the upper
  bound.
  This automaton can clearly be constructed on-the-fly in polynomial space.

  For the lower bound we  reduce from the problem of intersection of NFA.
  Let $\Aa_1,\dots,\Aa_n$ be NFA over the alphabet $\S$.
  We use processes $p_1,\dots,p_n$ where each $p_i$
  simulates $\Aa_i$.
  Process $p_1$ starts by guessing a letter $a$  of $\S$, making a
  transition on $a$
  and sending $a$ to $p_2$.
  Afterwards each  process $p_i$  receives a letter $a$ from $p_{i-1}$,
  makes a transition on $a$, then sends $a$ to $p_{i+1}$.
  Back again at $p_1$, the procedure restarts.
  Figure~\ref{fig:inter-auto-msc} shows the principle.

  Upon reaching a final state, $p_1$ can  send message \texttt{accept} to
  $p_2$ and then stop.
  If $p_i$  receives \texttt{accept} from  $p_{i-1}$ while being in a final state, it
  relays \texttt{accept} to $p_{i+1}$, and then stops.

  One can see that every "trace" of the "CFM" is "$\mb$-synchronizable",
  as every message is in its own "exchange". Moreover, the global-state $(\texttt{accept})_{p\in\Proc}$ is
  reachable if and only if the intersection of $\Aa_1,\dots,\Aa_n$ is non-empty.
\end{proof}

\begin{figure}
  \centering
  \scalebox{0.7}{\begin{tikzpicture}[overlay,yshift=-85.5pt,xshift=-33pt]%
    \node (e1) at (4.7,8.1) {\dots}; %
    \node (a1-alph) [fill=black,circle,minimum size=0.17cm,inner sep=0pt] at (3.83,6.79) {};
    \node (a1-alph-out) [minimum size=0pt] at (4.4,6.79) {};
    \draw[->] (a1-alph) edge[thick] (a1-alph-out) {};
    \node (e1) at (4.7,6.79) {\dots};
    \node (an1-alph-in) [minimum size=0pt] at (4.9,6.79) {};
    \node (an1-alph) [fill=black,circle,minimum size=0.17cm,inner sep=0pt] at (5.55,6.79) {};
    \draw[->] (an1-alph-in) edge[thick] (an1-alph) {};

    \node (a1-acc) [fill=black,circle,minimum size=0.17cm,inner sep=0pt] at (3.83,4.3) {};
    \node (a1-acc-out) [minimum size=0pt] at (4.4,4.3) {};
    \draw[->] (a1-acc) edge[thick] (a1-acc-out) {};
    \node (e1) at (4.7,4.3) {\dots};
    \node (an1-acc-in) [minimum size=0pt] at (4.9,4.3) {};
    \node (an1-acc) [fill=black,circle,minimum size=0.17cm,inner sep=0pt] at (5.55,4.3) {};
    \draw[->] (an1-acc-in) edge[thick] (an1-acc) {};

    \node[rotate=90] (e1) at (2.11,5.45) {\dots};
    \node[rotate=90] (e1) at (3.83,5.45) {\dots};
    \node[rotate=90] (e1) at (5.56,5.45) {\dots};
    \node[rotate=90] (e1) at (7.29,5.45) {\dots};

\end{tikzpicture}%
\begin{msc}[/msc/instance distance=0.1cm,label distance=2pt,font=\Large]{}
    \declinst{a0}{}{$p_1$}
    \declinst{a1}{}{$p_2$}
    \declinst{an1}{}{$p_{n-1}$}
    \declinst{an}{}{$p_n$}
    \mess{$a$}{a0}{a1}
    \nextlevel
    \nextlevel
    \mess{$a$}{an1}{an}
    \nextlevel
    \stop{a0}
    \stop{a1}
    \stop{an1}
    \stop{an}
    \nextlevel
    \startinst{a0}{}{}
    \startinst{a1}{}{}
    \startinst{an1}{}{}
    \startinst{an}{}{}
    \nextlevel
    \mess{\small\texttt{accept}}{a0}{a1}
    \nextlevel
    \nextlevel
    \mess{\small\texttt{accept}}{an1}{an}
\end{msc}}
  \caption{The "MSC" of a "trace" of the "CFM" for automata intersection.}
  \label{fig:inter-auto-msc}
\end{figure}

\section{Model-checking regular properties}
\label{sec:properties}

In this section we introduce a class of properties against which we can verify
"$\mb$-synchronizable"  "CFMs".
We look for regular properties $P$ over the alphabet $S \cup R \cup \bS$,
so we exploit the "marked sends" to refer (indirectly) to messages.
The model-checking problem we consider is the following:

\medskip

\noindent
\textsc{CFM-vs-regular property}\\
\textsc{Input:}  "$\mb$-synchronizable" "CFM" $\Aa$, regular property $P
  \subseteq (S \cup \bS \cup R)^*$.\\
\textsc{Output:}  Yes if for every "$\mb$-synchronous" "trace" $u \in \Ex_\mb(\Aa)$  we have $\mss(u) \in P$. 

\medskip

The  properties we consider are regular, "$R$-closed"
subsets of $(S \cup \bS \cup R)^*$:

\begin{mydefinition}[$R$-closed properties]
Let  $\equiv_R$ be the reflexive-transitive closure of the
    relation consisting of all pairs $(u \, a \, b \, v, u \,b\, a \, v)$ with $u,v
    \in (S \cup \bS \cup R)^*$,
    $a,b \in R$, and $a,b$ performed by distinct processes.
    A property $P \subseteq (S \cup \bS \cup R)^*$ is called
    ""$R$-closed"" if it is closed under $\equiv_R$ (i.e., for any $u
    \equiv_R v$ we have $u \in P$ iff $v \in P$).
\end{mydefinition}

As an example,
we can consider a system with a central process $c$ and a set of 
orbiting processes $p_1,\dots,p_n$.
The central process gives tasks to the orbiting processes,
and they  send back their results.
We can state a property expressing a round-based behavior for $c$: it sends tasks to orbiting
processes, and if a process $p_i$ does not send back to $c$
in the next round, it will not 
participate in further rounds anymore.
The opposite property consists of all sequences from
$A^*{S_c}^* c!p_i(m) {S_c}^* R^+ (\bigcup_{j\not=i}S_{p_j})^+ R^+
{S_c}^+R^* A^* p_i!c(m') A^*$  for some $i$ and $m,m'$, and $A =S \cup
\bar{S} \cup R$.
As the above property is "$R$-closed", its complement is too.

We will show that if the regular property is "$R$-closed" then the
model-checking problem stated above is \PSPACE-complete.
  Before that recall that both being "$\mb$-viable"
and being "$\mb$-synchronous" (assuming "$\mb$-viable") are non
regular properties.
However, it is not necessary to be able to express the above,  as we
will apply the property to "$\mb$-synchronous" "traces" of "CFM".
The next lemma is similar to Lemma~\ref{lem:automaton-ms}:

\begin{mylemma}[restrict-mb-sync]
  Let $P \subseteq (S \cup \bS \cup R)^*$ be regular and "$R$-closed".
  Then the set
  \[
    \text{Sync}(P)=\set{ v \in (S \cup \bS)^* \mid \exists u \text{
        "$\mb$-synchronous" } \text{ s.t. } v=\ms(u) \text{ and } \mss(u) \in P}
  \]
  is regular.
  If $P$ is given by an "$R$-diamond" NFA with $n$ states, then we can construct an NFA
  for $\text{Sync}(P)$ with $O(n^3\cdot 2^{|\Proc|})$ states. 
\end{mylemma}

\begin{proof}
  Let $P$ be given by an "$R$-diamond" NFA
  $\Pp=(L, S \cup \bS \cup R, \act{}_\Pp, \ell_0, F)$ with $n$ states.
  We may assume w.l.o.g.~that $\Pp$ contains no $\varepsilon$-transition.
  Consider the LTS with $\varepsilon$-transitions $\Pp_\sync$ obtained from Lemma~\ref{lem:automaton-ms}.
  Recall that this LTS has $O(n^3 \times 2^{|\Proc|})$ states.
  As NFA for $\text{Sync}(P)$, we take $\Pp_\sync$,
  with $(\ell_0,\es)$ as initial state,
  and $F \times 2^\Proc$ as final states.
\end{proof}

\begin{mytheorem}[Rclosed-model-checking]
The \textsc{CFM-vs-regular property} problem is \PSPACE-complete if the
property is  "$R$-closed".  
There exist properties that are not "$R$-closed" for which the problem
is undecidable.
\end{mytheorem}

\begin{proof}
  For the upper bound,
  consider an "$\mb$-synchronizable" "CFM" $\Aa = (\Aa_p)_{p\in \Proc}$ and
  an "$R$-closed" regular property $P \subseteq (S \cup \bS \cup R)^*$
  given by an NFA $\Pp$.
  Since $P$ is "$R$-closed",
  its complement $P^{co}$ is also "$R$-closed".
  As in the proof of Lemma~\ref{lem:aut-for-exchange},
  let $\Qq$ denote the asynchronous product $\bigasync_{p \in \Proc} \bar{\Aa}_p$,
  where each $\bar{\Aa}_p$ is the LTS obtained from $\Aa_p$ by
  adding a transition $\ell_p\xrightarrow{\bar{s}}_p \ell_p'$ for each transition $\ell_p\xrightarrow{s}_p \ell_p'$ with $s \in S$.
  Note that $\Qq$ is "$R$-diamond",
  so its language $Q = L(\Qq)$ is "$R$-closed".
  We derive that $Q \cap P^{co}$ is "$R$-closed".
  It is routinely checked that
  $(\Aa, \Pp)$ is a positive instance of \textsc{CFM-vs-regular property}
  iff
  the set $\text{Sync}(Q \cap P^{co})$,
  as defined in Lemma~\ref{lem:restrict-mb-sync},
  is empty.
  To derive the {\PSPACE} upper bound from this lemma,
  we still need to provide an "$R$-diamond" NFA for $Q \cap P^{co}$.
  This "$R$-diamond" NFA is simply the synchronous product of $\Qq$ and
  the minimal automaton of $P^{co}$.
  The latter is "$R$-diamond" since $P^{co}$ is "$R$-closed",
  and it can be constructed on-the-fly in polynomial space from $\Pp$.
  Now it suffices to check emptiness of the NFA for
  $\text{Sync}(Q \cap P^{co})$ from Lemma~\ref{lem:restrict-mb-sync}.
  The lower bound is again straightforward.

For the undecidability of model-checking a property that is not
"$R$-closed" we use a straightforward reduction from PCP.
Let $(u_i,v_i)_{i=1\dots k}$ be an instance of PCP over the binary alphabet $\{\mathtt{0}, \mathtt{1}\}$.
We can have three processes $p,U,V$ and process $p$ who sends, in
rounds, some pair $(u_i,v_i)$ to $U$ and $V$, resp.
That is, $p$ sends $u_i$ ($v_i$, resp.) letter by letter to $U$ ($V$,
resp.).
The processes $U$ and $V$ do nothing except receiving whatever $p$ sends to them.

There is a solution to the given PCP instance
iff
there is a "trace" consisting of a single fully "matched" "$\mb$-exchange"
where $U$ and $V$ perform the same receives in lock-step.
So we take as property $P$ the regular language
$P = (S \cup \bS \cup R)^* \setminus P^{co}$ where
$P^{co} = S^* \{U?p(\mathtt{0}) V?p(\mathtt{0}), U?p(\mathtt{1}) V?p(\mathtt{1})\}^*$.
\end{proof}

\subsection{Comparing $\ptp$ and $\mb$ semantics}
\label{sec:ptp_mb}
Given a protocol that was designed for $\ptp$ communication, it can be
useful to know whether the protocol can be also deployed under mailbox
communication.
We call this property mailbox-similarity.

\begin{mydefinition}[Mailbox-similarity]
  \AP
  A "$\ptp$-viable" sequence of actions $u$ is called ""mailbox-similar""
  if there exists some "$\mb$-viable" sequence $v$ such that $u \equiv v$.
  A "CFM" $\Aa$ is called \emph{mailbox-similar} if every "trace" from
  $\Ex_\ptp(\Aa)$   is "mailbox-similar".
\end{mydefinition}

For example, we have $p!q(a)\,r!q(b)\,q?p(a) \equiv r!q(b)\,p!q(a)\,q?p(a)$,
hence $r!q(b)\,p!q(a)\,q?p(a)$ is not "$\mb$-viable", but it is mailbox-similar.
Note that by Remark~\ref{rem:viable-vs-valid}, any "$\ptp$-viable" sequence $u$ is "mailbox-similar" iff
$\mscn(u)$ is "$\mb$-valid".
Unsurprisingly, as it is often the case under $\ptp$ semantics,
"mailbox-similarity" is undecidable without further restrictions:

\begin{figure}
  \centering
  \begin{minipage}[t]{0.49\textwidth}
    \centering
    \scalebox{0.7}{\input{figures/msc_mbx_sim_undec.tex}}
    \caption{Gadget to reduce reachability in "peer-to-peer semantics" to non-"mailbox-similarity".}
    \label{fig:msc-mbx-sim-undec}
  \end{minipage}
  \begin{minipage}[t]{0.45\textwidth}
      \centering
      \scalebox{0.7}{\input{figures/non-mb-sync-gadget.tex}}
      \caption{Gadget for non-"\mb-synchronizability".}
      \label{fig:non-sync-gadget}
  \end{minipage}
\end{figure}

\begin{mylemma}[undec-mb-sim]
  The question whether a given "CFM" is "mailbox-similar" is undecidable.
\end{mylemma}

\begin{proof}
  We show a reduction from the configuration reachability problem for "CFM" with "peer-to-peer semantics".
  This problem is well-known to be undecidable even when (a) there are only two processes and (b) the
  target configuration has empty buffers.
  Assume that we are given a "CFM" $\Aa$ with two processes $p$ and $q$ and let $(\ell_p, \ell_q)$ be a global state.
  We introduce a fresh message $\$$ and an additional process $r$ that
  does only $r?q(\$)$.
  In addition, we modify the processes $p$ and $q$ as follows.
  From the local state $\ell_p$,
  process $p$ may choose to move to a new local state and then perform
  $p!r(\$) p?q(\$) p!q(\$)$.
  From the local state $\ell_q$,
  the process $q$ may choose to move to a new local state and then perform $q!p(\$) q?p(\$) q!r(\$)$.
  Let us call $\Bb$ the resulting "CFM" with process $r$ and modified processes $p$ and $q$.
  We show that the configuration $((\ell_p, \ell_q), (\varepsilon, \varepsilon))$ is reachable in $\Tt_\ptp(\Aa)$ if,
  and only if,
  $\Bb$ is not "mailbox-similar".

  \smallskip

  Assume that $u$ is the label of an initial "execution" of $\Tt_\ptp(\Aa)$ ending in $((\ell_p, \ell_q), (\varepsilon, \varepsilon))$.
  Consider the "$\ptp$-viable" sequence
  $w = p!r(\$) q!p(\$) p?q(\$) p!q(\$) q?p(\$) q!r(\$) r?q(\$)$.
  The "MSC" of $w$ is depicted in Figure~\ref{fig:msc-mbx-sim-undec}.
  The sequence $v = u w$ is clearly a "$\ptp$-viable" "trace" of $\Bb$.
  Observe that there is a
  $(\lhb \cup <_\mb)$-cycle $p!r(\$) \lhb q!r(\$) <_\mb p!r(\$)$
  in $\mscn(v)$.
  Hence,
  $\mscn(v)$ is not "$\mb$-valid",
  which means that $v$ is not "mailbox-similar".

  \smallskip

  Conversely,
  assume that $v$ is a "trace" of $\Bb$ over $\ptp$ that is not "mailbox-similar".
  Since $v$ is "$\ptp$-viable" but not "$\mb$-viable", the process $r$
  necessarily moves in $v$.
  This is because any "$\ptp$-viable" sequence over two processes is
  also "$\mb$-viable".
  So $r$ performs $r?q(\$)$ in $v$.
  This in turn entails
  that the projection of $v$ on $p$ ends with $p!r(\$) p?q(\$) p!q(\$)$ and
  that the projection of $v$ on $q$ ends with $q!p(\$) q?p(\$) q!r(\$)$.
  It follows that $((\ell_p, \ell_q, i_r), (\varepsilon, \ldots, \varepsilon))$ is visited
  by an initial "execution" of $\Tt_\ptp(\Bb)$ whose label is equivalent to $v$.
  We derive that $((\ell_p, \ell_q), (\varepsilon, \varepsilon))$ is reachable in $\Tt_\ptp(\Aa)$.
\end{proof}

In the remainder of this section,
we show that "mailbox-similarity" becomes decidable if we assume
that the "CFM" is "$\mb$-synchronizable".
Recall that the latter means that every "trace" from
  $\Ex_\mb(\Aa)$ is "$\mb$-synchronizable".

The next lemma shows how to check that two positions in an
"$\mb$-synchronous" sequence $u$  are causally-ordered, i.e., there is some 
$(\lhb \cup <_\mb)$-path between these positions (as usual, this
refers to a path  between associated events in $\mscn(u)$).
We mark these positions  using a ``tagged'' alphabet
$\S = (S \cup \bS \cup R) \times \set{\circ, \bullet}$.

\begin{mylemma}[causality-check-auto-marked]
  We can construct an "$R$-diamond" automaton $\Dd$ with $O(|\Proc|)$ states
  over the alphabet $\S$ such that
  for every "$\mb$-synchronous" sequence $u \in \Act^*$ and every positions $i<j$ of $u$
  such that $u[i]$ and $u[j]$ are in $S$,
  there is a $(\lhb \cup <_\mb)$-path from $u[i]$ to $u[j]$
  iff
  $\Dd$ accepts the word $\mss(u)$ tagged by $\bullet$ at $i$ and $j$ and by $\circ$ elsewhere.
\end{mylemma}
\begin{proof}
  Recall that $\lehb=(<_\Proc\cup \msg)^*$ is the happens-before order.
  The automaton $\Dd$ will guess a $(<_\Proc\cup \msg \cup <_\mb)$-path
  from $u[i]$ to $u[j]$.
  It will actually use only send actions of $\mss(u)$, relying on the
  fact that $u$ is "$\mb$-synchronous".
  That is,
  $\Dd$ guesses a subsequence of positions $i_1< \dots < i_t$ of $u$,
  with each $u[i_k] \in S$,
  as described in the following.
  Let $i_0=i$ and $i_{t+1}=j$.
  We have three cases, and $\Dd$ guesses in which case we are:
  \begin{itemize}
  \item
    $u[i_k], u[i_{k+1}]$ are performed by the same process $p$.
    After $i_k$ the automaton $\Dd$ remembers
    the pair $(<_\Proc,p)$ until it guesses $i_{k+1}$.
  \item
    $u[i_k], u[i_{k+1}]$ are both sends to the same process $p$,
    and $u[i_k]$ is "matched".
    After $i_k$ the automaton $\Dd$ remembers
    $(\mbord,p)$ until it guesses $i_{k+1}$.
  \item
    $u[i_k]$ is "matched",
    its receive $u[h]$ is performed by the same process $p$
    as $u[i_{k+1}]$, and $h < i_{k+1}$.
    After $i_k$ the automaton $\Dd$ remembers
    the pair $(\msg,S,p)$.
    After the next receive action,
    $\Dd$ changes its state to $(\msg,R,p)$
    until it guesses $i_{k+1}$.
    The assumption that $u$ is "$\mb$-synchronous" guarantees that
    the receive $u[h]$ "matched" with $u[i_k]$ has already occurred
    when $\Dd$ guesses $i_{k+1}$.
  \end{itemize}
  By construction, if $\Dd$ accepts $\mss(u)$,
  then we have a $(\lhb \cup <_\mb)$-path
  from $u[i]$ to $u[j]$,
  with $i<j$ the two positions tagged by $\bullet$ in $\mss(u)$.

  For the left-to-right implication,
  assume that $u[i]$ and $u[j]$ are in $S$ and that
  we have a $(\lhb \cup <_\mb)$-path from $u[i]$ to $u[j]$.
  This path is a
  sequence $i=i_0 < i_1 \dots < i_t <i_{t+1}=j$ of positions of $u$,
  such that each pair of consecutive indices  is related by $<_\Proc$, $<_\mb$
  or $\msg$.
  Moreover,
  we may assume w.l.o.g.~that there are no two consecutive $<_\Proc$-arcs on this path.
  If the path contains only $<_\Proc$ and $<_\mb$-arcs,
  then $\Dd$ applies one of the first two rules above.
  Consider now a $\msg$-arc $(u[i_k], u[i_{k+1}])$.
  As $u[i_{k+1}]$ is a receive,
  we get that $u[i_{k+1}] <_\Proc u[i_{k+2}]$.
  Moreover,
  $u[i_{k+2}]$ is a send since there are no two consecutive $<_\Proc$-arcs on the path.
  So $\Dd$ can apply the third rule to go from $i_k$ to $i_{k+2}$.
  We get that $\Dd$ accepts the word $\mss(u)$ tagged by $\bullet$ at
  $i$ and $j$ and by $\circ$ elsewhere.
    The number of states of $\Dd$ is  $4*|\Proc| +2$
  ($2$ for initial/final state).
\end{proof}

\begin{myremark}[viable-vs-similar]
  Before we state the next lemma we note that if $u=v\, r$ with $r \in R$ is
"mailbox-similar" then there exists some $v'$ such that $u \myequiv v' \, r$ and
$v'\, r$ is "$\mb$-viable". Note also that if $u=v\, r$ is "mailbox-similar"  and
$v$ is 
"$\mb$-viable" then $u$ is not necessarily "$\mb$-viable" (see also Remark~\ref{rem:non-viable}).
\end{myremark}

The next lemma states an inductive property of "mailbox-similar" sequences. 
It will be also used in Section~\ref{sec:atomic}.

\begin{mylemma}[mb-viable]
  Let $u = v \, r$ be a "$\ptp$-viable" sequence with $r\in R$ such that
  $v$ is "$\mb$-viable". 
  Let $q$ be the process performing the receive $r$.
  Then $u$ is "mailbox-similar"  if and only if there is no non-empty
  $(\lhb\cup <_\mb)$-path from $v[i]$ to $v[j]$ for some $i<j$ such that $v[i]$
  is an "unmatched" send to $q$ and $v[j]$ is the send matching $r$ in $u$.
\end{mylemma}
\begin{proof}
  First, we observe that every arc present  in $\mscn(v)$ is also
  present in $\mscn(u)$. %
  For the left-to-right direction, assume that there is some $(\lhb \cup
  <_\mb)$-path from $v[i]$ to $v[j]$ for $i,j$ as in the statement.
  Since $v[j]$ becomes "matched" in $u$ we have an $<_\mb$-arc from
  (the event corresponding to) $v[j]$
  to (the event corresponding to) $v[i]$, so this  creates a cycle for the relation
  $(\lhb \cup <_\mb)$ in $\mscn(u)$, hence $\mscn(u)$ is not "$\mb$-valid".
  By Remark~\ref{rem:viable-vs-valid}, the sequence $u$ is not "mailbox-similar".

  For the right-to-left direction, if $u$ is not 
  "mailbox-similar", then
  $\mscn(u)$ contains a $(\lhb \cup <_\mb)$-cycle.
  The only arcs we added by doing $r$ 
  are  $<_\mb$-arcs from $v[j]$ to every $v[i]$ corresponding to an
  "unmatched" send to $q$,
  and the message arc from $v[j]$ to $r$.
  The cycle can arise only if 
  we have a $(\lhb \cup <_\mb)$-path from some $v[i]$ to $v[j]$ with
  $i<j$ and such that $v[i]$ is an "unmatched" send to $q$.
\end{proof}

\begin{mylemma}[mailbox-similarity-check-auto]
  For any receive action $r \in R$,
  we can construct an "$R$-diamond" automaton $\Pp_r$ with $O(|\Proc|)$ states
  over the alphabet $(S \cup \bS \cup R)$ such that
  for every "$\mb$-synchronous" sequence $u$,
  it holds that
  $u \, r$ is "$\ptp$-viable" and not "mailbox-similar"
  iff  $\Pp_r$ accepts
  $\mss(u)$.
\end{mylemma}

\begin{proof}
  Consider a receive action $r = q?p(m)$.
  Let $W_r$ denote the set of words $w \in \S^*$ such that
  $w$ contains exactly two positions $i < j$ tagged by $\bullet$,
  $w[i]$ is an "unmatched" send to $q$,
  $w[j]$ is $\bar{p!q(m)}$,
  and no $w[h]$ with $h < j$ is an "unmatched" send from $p$ to $q$.
  It is easily seen that $W_r$ is recognized by an "$R$-diamond" NFA $\Ww_r$ with three states.
  Let $\Ee_r$ denote the synchronous product of $\Ww_r$ and
  the "$R$-diamond" automaton $\Dd$ from Lemma~\ref{lem:causality-check-auto-marked}.
  The desired automaton $\Pp_r$ is obtained from $\Ee_r$ by untagging it,
  that is,
  by replacing each tagged action $(a, t) \in \S$ by $a$.
  As $\Ee_r$ is "$R$-diamond", so is $\Pp_r$.
  Let us show that $\Pp_r$ satisfies the condition of the lemma.%
  We assume,
  for the remainder of the proof,
  that $v$ is an "$\mb$-synchronous" sequence.

  \smallskip

  Suppose that $u = v \, q?p(m)$ is "$\ptp$-viable" and not "mailbox-similar".
  Let $j$ denote the position in $v$ of the send matching $q?p(m)$.
  Note that $v[j]$ is a send $p!q(m)$ that is "unmatched" in $v$.
  Since $u$ is "$\ptp$-viable",
  no $v[h]$ with $h < j$ is an "unmatched" send from $p$ to $q$.
  Moreover,
  by Lemma~\ref{lem:mb-viable},
  there is a non-empty
  $(\lhb\cup <_\mb)$-path from some $v[i]$ with $i < j$ to $v[j]$ such that
  $v[i]$ is an "unmatched" send to $q$.
  Let $w$ denote the word $\mss(v)$ tagged by $\bullet$ at $i$ and $j$ and by $\circ$ elsewhere.
  By construction,
  we have $w \in W_r$ and $w \in L(\Dd)$,
  hence,
  $w \in L(\Ee_r)$.
  It follows that the untagged word $\mss(v)$ is in $L(\Pp_r)$.

  \smallskip

  Conversely,
  if $\mss(v) \in L(\Pp_r)$ then $\mss(v)$ is obtained by untagging some $w$ in $L(\Ee_r) = W_r \cap L(\Dd)$.
  We derive from the definition of $W_r$ and the property satisfied by $\Dd$ that
  there exist two positions $i < j$ in $v$ such that,
  on the one hand,
  $v[i]$ is an "unmatched" send to $q$,
  $v[j]$ is $\bar{p!q(m)}$,
  and no $v[h]$ with $h < j$ is an "unmatched" send from $p$ to $q$,
  and on the other hand,
  there is a $(\lhb \cup <_\mb)$-path from $v[i]$ to $v[j]$.
  It follows that $v \, q?p(m)$ is "$\ptp$-viable" and,
  by Lemma~\ref{lem:mb-viable},
  that $v \, q?p(m)$ is not "mailbox-similar".
\end{proof}

We derive from the previous lemma that
"mailbox-similarity" can be solved in {\PSPACE}
for "$\mb$-synchronizable" "CFMs".

\begin{mytheorem}[mailbox-similarity-pspace]
  The question whether a given "$\mb$-synchronizable" "CFM" is "mailbox-similar"
  is {\PSPACE}-complete.
\end{mytheorem}
\begin{proof}
  For the upper bound,
  consider an "$\mb$-synchronizable" "CFM" $\Aa = (\Aa_p)_{p\in \Proc}$.
  We first observe that
  $\Aa$ is not "mailbox-similar" iff
  there exists $r \in R$ and an "$\mb$-synchronous" sequence $u$ such that
  $u \, r$ is a "trace" in $\Ex_\ptp(\Aa)$ and $u \, r$ is not "mailbox-similar".
  The ``if'' direction is trivial.
  The ``only if'' direction is shown by taking a
  non-"mailbox-similar" "trace" $u=va$ in $\Ex_\ptp(\Aa)$ of minimal length.
  The last action $a$ of $u$ cannot be a send because $v$ is "mailbox-similar"
  by minimality, so $u$ would be "mailbox-similar" too, if $a$ were a send.

  As in the proofs of Lemma~\ref{lem:aut-for-exchange} and
  Theorem~\ref{thm:Rclosed-model-checking},
  let $\Qq$ denote the asynchronous product $\bigasync_{p \in \Proc}
  \bar{\Aa}_p$, %
  where each $\bar{\Aa}_p$ is the LTS obtained from $\Aa_p$ by
  adding a transition $\ell_p\xrightarrow{\bar{s}}_p \ell_p'$ for each transition $\ell_p\xrightarrow{s}_p \ell_p'$ with $s \in S$.
  Moreover,
  given $r \in R$,
  let us define the language $Q_r$ as the right derivative
  $Q_r = \{w \in (S \cup \bS \cup R)^* \mid w \, r \in L(\Qq)\}$.
  Note that $\Qq$ is "$R$-diamond",
  so $Q_r$ is "$R$-closed".
  Let $\Pp_r$ denote the "$R$-diamond" automaton obtained from Lemma~\ref{lem:causality-check-auto-marked},
  and let $P_r$ denote its "$R$-closed" language $P_r = L(\Pp_r)$.
  It is routinely checked that
  $\Aa$ is "mailbox-similar"
  iff
  for every $r \in R$,
  the set $\text{Sync}(Q_r \cap P_r)$,
  as defined in Lemma~\ref{lem:restrict-mb-sync},
  is empty.
  To derive the {\PSPACE} upper bound from this lemma,
  we provide,
  as "$R$-diamond" NFA for $Q_r \cap P_r$,
  the synchronous product of $\Qq_r$ and $\Pp_r$,
  where $\Qq_r$ is obtained from $\Qq$ by
  considering as final those "global states" $g$ such that
  there is a transition $g \xrightarrow{r} g'$ in $\Qq$.
  Note that $\Qq_r$ and $\Pp_r$ can both be constructed on-the-fly in polynomial space.
  Now it suffices to check emptiness of the NFA for
  $\text{Sync}(Q_r \cap P_r)$ from Lemma~\ref{lem:mailbox-similarity-check-auto}.

  \medskip

  For the lower bound, we use the same reduction as in Theorem~\ref{thm:reach-for-mb-sync},
  and if we reach $(\texttt{accept})_{p\in\Proc}$, we use two other processes to do a
  non-"mailbox-similar" gadget (for example
  Figure~\ref{fig:msc-mbx-sim-undec}).
  This way, the "CFM" is "mailbox-similar" if and only
  if the intersection of the automata $\Aa_1,\dots,\Aa_n$ is empty.
\end{proof}

\section{Checking $\mb$-synchronizability}
\label{sec:sync}
In this section we show our main result, namely an algorithm to know if a "CFM" is
"$\mb$-synchronizable".
As a side result we obtain optimal complexity
bounds for some problems considered
in~\cite{digiusto20fossacs,GiustoLL23}.

The high-level schema of the algorithm is to look for a minimal
witness for non-"$\mb$-synchronizability".
This amounts to searching for an "$\mb$-synchronous" "trace" that violates
"$\mb$-synchronizability" after adding one (receive) action.
Of course, we need Theorem~\ref{thm:reach-for-mb-sync} to
guarantee that the "$\mb$-synchronous" "trace" is executable.
In addition, we have to detect the violation of
"$\mb$-synchronizability", and for this we need to determine if an
"exchange" is non-decomposable into smaller "exchanges".
Section~\ref{sec:atomic} shows automata for  non-decomposable
"exchanges", and in Section~\ref{sec:algorithm} we present the
algorithm that finds  minimal witnesses.

\subsection{Automata for atomic "exchanges"}\label{sec:atomic}

\begin{figure}
  \centering
    \centering
      \scalebox{0.7}{\input{figures/two_scc_order_example.tex}}
    \caption{"MSC" of 
    $u=p_2!p_1(a)\,p_1!p_2(b)$ $p_1?p_2(a)\,
      p_2?p_1(b)\;p_3!p_2(c)$, with
    the two SCCs of its "communication graph". %
    Note that $1 \preceq_\mb^u2$,
    but neither $1 \preceq_\ptp^u2$ nor $2 \preceq_\ptp^u1$ holds.}
    \label{fig:skel_ord}
\end{figure}

In this section we consider sequences of
 actions  that cannot be split into smaller pieces without separating
 messages~\cite{GenestMSZ06,GiustoLL23}. 
We  introduce these notions for arbitrary  "many-to-one"
"process networks" $\Nn$.
Later we will fix $\Nn=\mb$ since
 reachability over synchronizable sequences is decidable in this
 setting.

\begin{mydefinition}[Atomic sequences]
  An "$\Nn$-viable" sequence $u \in \Act^*$ is ""$\Nn$-atomic"" (or
  \emph{atomic} for short)
   if $u \equiv v \astn w$ with $v,w$ both "$\Nn$-viable" implies that one of
   $v,w$ is empty.
\end{mydefinition}

To check atomicity we can use a graph criterium introduced already
in~\cite{HL00} (see also \cite{GenestMSZ06}), that is similar to the
notion of conflict graph used in \cite{bouajjani2018completeness}:

\begin{mydefinition}[Communication graph]
  \AP Let $u$ be an "$\Nn$-viable" sequence, and $\Msc=\mscn(u)$.
  The ""$\Nn$-communication graph"" of $u$ is the directed graph $\G_\Nn(u)=(V,E)$
  where $V$ is the set of all events of $\Msc$ and the edges are defined
  by $ ( e,e')\in E $ if $e <_\Proc e'$ or $e <_\Nn e'$ or
  $\{(e,e'),(e',e) \} \cap \msg \not=\emptyset$.
\end{mydefinition}

  The right part of Figure~\ref{fig:labeling} shows (partly) the   "communication graph" of the "MSC" in the left part.
  The cycle witnesses that the "MSC" is "$\Nn$-atomic" for $\Nn \in
  \set{\mb, \ptp}$, according to the next lemma.

\begin{mylemma}[indivisibility-from-inter]
  Let $u \in \Act^*$ be a "$\Nn$-viable" sequence
  and $\G_\Nn(u)$ the "$\Nn$-communication graph" of $\mscn(u)$.
  Then $u$ is "$\Nn$-atomic"
  if and only if $\G_\Nn(u)$ is strongly connected.
\end{mylemma}

\begin{proof}
  
  For the left-to-right implication let us suppose that $u$ is "$\Nn$-atomic", but there
  is more than one strongly connected component (SCC for short) in
  $\G_\Nn(u)$.
  Assume that the SCCs of $\G_\Nn(u)$ are $C_1,\dots, C_k$, sorted in
  some  topological order.
  First we claim that  for every SCC $C_i$, the restriction $\Msc_i:=\Msc_{|C_i}$
  of $\mscn(u)$ to $C_i$ is an "$\Nn$-valid" $\mscn$.  
  First  note that any two events that form a message 
  in $\mscn(u)$ are in the same $\Msc_i$, for some $i$. As the "process order"
  $<_\Proc$ of $\Msc_i$ is
  inherited from $\mscn(u)$, and the network order $<_\Nn$ is constructed from the two other orders,
  $\Msc_i$ is "$\Nn$-valid".
  Consider for each $i$ a linearization $u_i$ of the partial order
  $(\lehb \cup \leqb)^*$ of $\Msc_i$, and
  recall that each $u_i$ is "$\Nn$-viable".
  We claim now that $u_1 \astn \dots \astn u_k$ is defined.
  Otherwise there would be a "buffer" $\bufelt$ such that 
  there is an "unmatched" send to $\bufelt$ in $u_i$ and a "matched" send to
  $\bufelt$ in $u_j$, for some $i<j$. This would
  mean that there is an edge from $C_j$ to $C_i$ in $\G_\Nn(u)$ (caused by $<_\Nn$),
  which contradicts the topological order. Since $u \equiv u_1 \astn
  \dots \astn u_k$, we obtain a contradiction to $u$ being "$\Nn$-atomic".

  Now we show the right-to-left implication.
  We suppose that $\G_\Nn(u)$ is strongly connected,
  but $u$ is not "$\Nn$-atomic".
  Then there exist $v$ and $w$ non-empty "$\Nn$-viable" sequences such that
  $u \equiv v \astn w$. We then have that, in $\mscn(u)$, there is no $\msg$-arc between events of 
  $v$ and $w$, and there is neither a $<_\Proc$ nor a $<_\Nn$-arc from an event of $w$  to an event of $v$. Therefore, in $\G_\Nn(u)$ there is no path from any event of $w$ to 
  any event of $v$, so $\G_\Nn(u)$ is not strongly connected. Contradiction.
\end{proof}

From Lemma~\ref{lem:indivisibility-from-inter} we can infer a decomposition of any
"trace" in "atomic" subsequences that is unique up to permuting
adjacent "atomic" sequences that are not ordered in the sense of the
next definition:

\begin{mydefinition}[Skeleton][order-atoms]
  Let $u$ be a "$\Nn$-viable" sequence with $\Msc = \mscn(u)$ 
  and $\G_\Nn(u)$ be the "$\Nn$-communication graph" of $\Msc$.
  Fix some arbitrary topological indexing $\set{1,\dots,n}$  of the SCCs of $\G_\Nn(u)$.
  We define the \emph{skeleton} of $u$ as $\skel(u)
  = (\set{1,\dots,n},\preceq_\Nn^u)$, where $\preceq_\Nn^u$ is the
  partial order induced by setting $i\prec_\Nn^u j$ for $1 \le i < j
  \le n$  if there is some $<_\Proc$-arc or
  some $\mb$-arc in
  $\G_\Nn(u)$ from the SCC with index $i$ to the SCC with index $j$.
\end{mydefinition}

\begin{myremark}[order-on-sequence]
  Assume that $u = u_1 \ast_\Nn \dots \ast_\Nn u_n$ where each $u_i$
  is "$\Nn$-atomic" and non-empty,
  and we index the SCCs according to the order of the $u_i$.
  Then we obtain $\skel(u)= (\set{1,\dots,n},\preceq_\Nn^u)$ with $i \prec_\Nn^u
  j$ if either both $u_i$ and $u_j$ contain some 
  actions on the same process; or they both contain some send to
  the same  "buffer", with the one in $u_i$ being "matched".
  See Figure~\ref{fig:skel_ord} for an example.
\end{myremark}

\begin{mylemma}[dec-into-atomic]
  Let $u$ be an "$\Nn$-viable" sequence.
  Then there exist some "$\Nn$-atomic" non-empty sequences $u_1,\dots, u_k$ such that
  $u \equiv u_1 \astn \dots \astn u_k$.
  Such a decomposition into  "$\Nn$-atomic" non-empty sequences is
  unique up to the partial order  $\preceq_\Nn^u$ of $\skel(u)$. 
\end{mylemma}

\begin{proof}
  Let $\G_\Nn(u)$ be the  "$\Nn$-communication graph" of  $\mscn(u)$.
  By Lemma~\ref{lem:indivisibility-from-inter} the SCCs $C_1,\dots,
  C_k$ of $\G_\Nn(u)$ induce  "$\Nn$-atomic" subsequences $u_1,\dots, u_k$
  with $u \equiv u_1 \astn \dots \astn u_k$.
  Conversely, if $u \equiv u' \astn v \astn u''$, with $v$ non-empty and
  "$\Nn$-atomic", then $v$ induces an SCC of $\G_\Nn(u)$.
  This is due to  Lemma~\ref{lem:indivisibility-from-inter}, and to the
  fact that our product $\astn$ prevents backward $<_\Nn$-arcs in
  $\G_\Nn(u)$ from $v$ to $u'$, or from $u''$ to $v$.
  Note also that the partial order on the SCCs of $\G_\Nn(u)$ is generated by
  $<_\Proc$-arcs or $<_\Nn$-arcs, which yields the second statement. 
\end{proof}

Throughout the remaining of this section we fix $\Nn=\mb$.
We will show now a simple,
automaton-compatible condition to certify that an 
"ms-sequence" $v=\ms(u)$ 
corresponds to an "$\mb$-atomic" "exchange" $u$.
First we note that, in order for the "communication
graph" $\G_\mb(u)$ to be strongly connected, there must exist for every  process
$p$ that is active in $u$ some path from the last action of $p$ to the first action of
$p$ (if there are at least two actions of $p$ in $u$).
A process $p$ is called ""active"" in $u$ if there is at least some action
performed by $p$ in $u$ (resp., if $v=\ms(u)$ contains either a send
performed by $p$, or a "matched" send to $p$).
We look for such a path for every "active" process and
then we need to connect all such paths together. 

\begin{figure}[t]
  \centering
  \begin{subfigure}[c]{0.4\textwidth}
      \scalebox{0.8}{\input{figures//msc_running_example_atomic.tex}}
  \end{subfigure}
  \begin{subfigure}[c]{0.5\textwidth}
      \centering
      \scalebox{0.7}{\begin{tikzpicture}[
    > = stealth, %
    auto,
    node distance = 2cm, %
    thick, %
    inner sep=1pt,
    minimum size = 0.05cm,
    every node/.style={draw,circle},
    font=\Large
]

\tikzstyle{scc2} = [color = black,opacity=0.15]
\tikzstyle{alert} = [color=red!70]
\begin{scope}[node distance=1.5cm]
    \node[draw=none] (a) at (1,1) {};
    \node (r3) [left=of a,alert] {$r_d$};       \node (r1) [right=of a] {$r_b$};
    \node (s0) [above=of a] {$s_a$};      \node (r2) [below=of a] {$r_c$};
    \node (r0) [above left=of a,alert] {$r_a$}; \node (s1) [above right=of a] {$s_b$};
    \node (s3) [below left=of a] {$s_d$}; \node (s2) [below right=of a] {$s_c$};
    \node[scc2] (s4) [below left= 6pt and 6pt of s3] {$s_e$};
    \node[scc2] (r4) [below left= 8pt and 15pt of r3] {$r_e$};
  \end{scope}

  \tikzstyle{label}=[color=Green4!90]

  \begin{scope}[node distance=2.1cm,inner sep=2pt, minimum size=0.2cm,label,draw=white]
    \node[draw=none] (a') at (1,1) {};
    \node[draw=none] (r3') [left=of a] {0};       \node[draw=none] (r1') [right=of a] {2};
    \node[draw=none] (s0') [above=of a] {};      \node[draw=none] (r2') [below=of a] {1};
    \node[draw=none] (r0') [above left=of a] {3}; \node[draw=none] (s1') [above right=of a] {};
    \node[draw=none] (s3') [below left=of a,draw=none] {}; \node[draw=none] (s2') [below right=of a] {};
  \end{scope}

\path[<->] (s0) edge[bend right=13] (r0);
\path[<->] (s1) edge[bend left=13] (r1);
\path[<->] (s2) edge[bend left=13] (r2);
\path[<->] (s3) edge[bend left=13] (r3);
\path[->] (r0) edge[bend right=13] (r3);
\path[->] (s3) edge[bend right=13] (r2);
\path[->] (s1) edge[bend right=13] (s0);
\path[->] (s2) edge[bend right=13] (r1);
\path[->] (s0) edge[]              (s3);
\path[->,scc2] (r3) edge[]              (r4);
\path[->,scc2] (r2) edge[bend left=5]   (s4);
\path[<->,scc2] (s4) edge[bend left=13] (r4);

\draw[->,label,shorten <=0.3cm,shorten >=0.2cm] (r3') arc (180:270:2.43cm);
\draw[->,label,shorten <=0.3cm,shorten >=0.2cm] (r2') arc (-90:0:2.43cm);
\draw[->,label,shorten <=0.3cm,shorten >=0.2cm] (r1') arc (0:135:2.43cm);

\end{tikzpicture}}
      \scalebox{0.8}{$\overset{\text{\normalsize\color{Green4!90}2}}{p_2!p_3(b)}\,,\,%
        \overset{\text{\normalsize\color{Green4!90}3}}{p_2!p_1(a)}\,,\,%
        \overset{\text{\normalsize\color{Green4!90}0}}{p_2!p_1(d)}\,,\,%
        \overset{\text{\normalsize\color{Green4!90}1}}{p_3!p_2(c)}$}
  \end{subfigure}
  \caption{A "well-labeling" of the "ms-sequence" bottom right,  
  witnessing a path in the communication graph of the "MSC" left, from the
   last to the first event of process $p_1$. 
     The $s_m$ and $r_m$ vertices of the communication graph
   correspond respectively to the send and receive of message $m$.}
  \label{fig:labeling}
\end{figure}
\AP Let $u \in Act^*$ be an "$\mb$-exchange".
For some suitable integer $n$ we define a labeling
of $v=\ms(u)$ as an injective
mapping $\pi : \set{0,\dots,n-1} \rightarrow \set{1,\dots,|v|}$ where
$\pi(i) = j$ means that position $j$ of $v$ is labeled by $i$.
We say that
$\pi$ is a ""well-labeling"" of $v$ (of size $n$) if, for every $0\leq i < n$:
\begin{itemize}
\item either $\pi(i)< \pi(i+1)$ and, for some process  $p$:\\
   \begin{tabular}{l}
        $v[\pi(i)]$ and 
          $v[\pi(i+1)]$ are both sends by $p$, or\\
        $v[\pi(i)]$ and  $v[\pi(i+1)]$ are both  sends to $p$, with $v[\pi(i)]$ "matched"
   \end{tabular}
   (""direct arc"")
\item or $v[\pi(i)]$ is a send by $p$ and 
        $v[\pi(i+1)]$ is a "matched" send to $p$ (""indirect arc"").  
\end{itemize}

\noindent 
An example of such labeling  is shown in~\cref{fig:labeling}.
Informally, one can see the two types of arcs between positions of $v$ as:

\begin{itemize}
  \item A "direct arc" between two sends corresponds to the "process order"
    $\leq_\Proc$ or the  "mailbox order" $\leq_\mb$ in $\mscn(u)$.
 For example, we have a "direct arc" from position 2 to 3 in~\cref{fig:labeling}.
  \item An "indirect arc" between two sends stems from composing edges of the
    "communication graph" $\G_\mb(u)$ that involve a receive event. 
  An "indirect  arc" is specific to "$\mb$-exchanges": in  $\G_\mb(u)$
  we can go from  the event of $v[i]$ 
  to the receive associated with the event of $v[j]$ (since $u$ is an
  "$\mb$-exchange" this receive is after $v[i]$), and then follow the message edge
  backwards to the event of $v[j]$.
 For example, we have an "indirect arc" from position 1 to 2 in
 \cref{fig:labeling}.
 \end{itemize}

\begin{mylemma}[well-label]
  Let $u$ be an "$\mb$-exchange" with $\Msc=\mscn(u)$, and $v=\ms(u)$.
  There is a path in the "communication graph" $\G_\mb(u)$ 
 from the event of $\Msc$ corresponding to $v[i]$ to the event
 corresponding to $v[j]$ if and only if there is a "well-labeling" of
  $v$ starting at $i$ and ending at $j$.
\end{mylemma}

\begin{proof}
   For the right-to-left direction, let $\pi$ be a "well-labeling" of $v$
  starting at $i$ and ending at $j$. As $\pi$ is a "well-labeling", 
  there is a path in $\G_\mb(u)$ from the event corresponding to
  $v[\pi(k)]$ to the one of  $v[\pi(k+1)]$, for every $k$ in the domain
  of $\pi$. Each such path is either a direct edge, or consists of two
  edges, as explained before the statement of the lemma in the main
  body.

  For the left-to-right direction, we suppose there is a path $\Pi$ in
  $\G_\mb(u)$ 
  from the event of $v[i]$ to the event of $v[j]$.
  We construct
  a labeling $\pi$ of $v$ that starts at $i$ and ends at $j$, by
  labelling the positions of $v$ that correspond to the events of $\Pi$ with
  their respective rank on $\Pi$.
  Suppose that $n$ positions are labeled and  let $0\leq k< n$. We
  show the existence of an arc from $\pi(k)$ to $\pi(k+1)$, which is
  either direct or indirect.
  There are three cases:
  \begin{itemize}
    \item There is no receive between
      the event of $v[\pi(k)]$ and the one of $v[\pi(k+1)]$ on $\Pi$.
      Thus $v[\pi(k)]$, $v[\pi(k+1)]$ are consecutive on $\Pi$ and
      are either ordered by $<_\Proc$ or by $<_\mb$.
      This gives a  "direct arc" from $\pi(k)$ to $\pi(k+1)$.
    \item Between the event of
      $v[\pi(k)]$ and the one of  $v[\pi(k+1)]$ we see on $\Pi$ the receive matching
      $v[\pi(k)]$ before  the receive  matching $v[\pi(k+1)]$.
    Note that both receives must be on the same process (as all
    receives between  $v[\pi(k)]$ and $v[\pi(k+1)]$), so they are ordered by
    $<_\Proc$.
    Thus, the events of $v[\pi(k)]$ and $v[\pi(k+1)]$, respectively,
    are ordered by $<_\mb$.
     This gives a  "direct arc" from $\pi(k)$ to $\pi(k+1)$.
    \item Between the event of
      $v[\pi(k)]$ and the one of  $v[\pi(k+1)]$ we have on $\Pi$ the receive matching
      the event of $v[\pi(k+1)]$ on the same process as the event of $v[\pi(k)]$.
      This gives an "indirect arc" from $\pi(k)$ to $\pi(k+1)$.
      \qedhere
  \end{itemize}
\end{proof}
\begin{myremark}
  In Lemma~\ref{lem:well-label}, we only talk about send actions.
  If we are interested in a path in $\G_\mb(u)$ to a receive action, we just need
  to exhibit the path to its corresponding send action.
\end{myremark}

We can infer a bound on the size of "well-labelings", using the pigeonhole principle on the
"direct arcs" and "indirect arcs" going through each process.

\begin{mylemma}[well-label-size]
  Let $u \in \Act^*$ be an "$\mb$-exchange" and $v=\ms(u)$.
  If there is a path in the "communication graph" $\G_\mb(u)$ between 
  $v[i]$ and $v[j]$ then there is a "well-labeling" of $\ms(u)$
  starting at position $i$ and ending at position $j$ of size at most
  $|\Proc|^2+|\Proc|$. 
\end{mylemma}
\begin{proof}
  To show this bound, we first bound the number of "indirect arcs",
   and then show that between two "indirect
  arcs" we  have a bounded number of positions.

  Suppose that $\pi$ is a minimal "well-labeling" starting in
  $i$, ending in $j$, and with more than
  $|\Proc|$ "indirect arcs" between consecutive positions.
  So there are at least two "indirect arcs" on $\pi$ involving the same
  process. 
  Let $k < k'$ be 
  be such that both $\pi[k],\pi[k+1]$ and $\pi[k'],\pi[k'+1]$ two
  indices are "indirect arcs" involving process $p$. 
  So we know that $v[\pi(k+1)]$ and 
  $v[\pi(k'+1)]$ are both of type $S_{\rightarrow p}$.
  Thus we also have an "indirect arc" from $v[\pi(k)] $ to
  $v[\pi(k'+1)]$. 
  We could then shorten the size of the "well-labeling", which
  contradicts the minimality of $\pi$.

  Now let us suppose that we have a minimal
  "well-labeling" with at most $|\Proc|$ "indirect arcs" and of size
  larger than $|\Proc|^2+|\Proc|$. This means there exist two indices
  $k < k'$ of $\pi$, such that  $k'-k > |\Proc|$ and such that all arcs
  in $\pi[k],\dots, \pi[k']$ are "direct arcs".
  In particular, $\pi[k]<\pi[k+1]< \cdots < \pi[k']$.
  So we have
at least two "direct arcs" $\pi[\ell_1],\pi[\ell_1+1]$ and
$\pi[\ell_2],\pi[\ell_2+1]$ involving the
same process $p$ and such that $k \le \ell_1 < \ell_2 \le k'$.
Each of the two "direct arcs" is of type "process order" or "mailbox
order".
One can check that in all combinations there is a "direct arc" from $\pi[\ell_1]$ to
$\pi[\ell_2]$, so we can obtain a  smaller "well-labeling",  contradicting minimality.
\end{proof}

We construct now two kinds of automata, both working on "ms-sequences" $v=\ms(u)$.
Automaton $\Bb_p$ will check for a  process $p$ that
is "active" in $u$, that  all
actions performed by $p$  are on a cycle in $\G_\mb(u)$. %
Automaton $\Bb_{all}$ will check that all actions of "active" processes in
$u$ appear together on a cycle in  $\G_\mb(u)$, by looking for a cycle going through all
active processes at least once.
In both cases, we construct the automaton as follows.
The states are lists of bounded length consisting  of pairs of sends ("marked" or not) and
\emph{timestamps}, and representing  "well-labelings".
The initial state is the empty
list. When an element of the "ms-sequence" is read, the automaton can non-deterministically
choose to add it somewhere in the list,  recording  when it
was added (i.e., how many elements were added before it).
The final states are those in which the list with the order of
insertions corresponds to a "well-labeling" of the "ms-sequence". We
can slightly modify 
this automaton to obtain $\Bb_p$ and $\Bb_{all}$, resp.
Below, if the first or last action on some process is a receive,
then we add its matching send to the list.
\begin{itemize}
  \item 
  For $\Bb_p$ the lists are of length at
  most $|\Proc|^2+|\Proc|$ (by Lemma~\ref{lem:well-label-size}).
  The final states should also require that
  the first element of the list is the last action of $p$ and the last element of
  the list is the first action of $p$.
  \item 
  For $\Bb_{all}$ the lists are of length at
  most $|\Proc|\cdot( |\Proc|^2+|\Proc|)$.
  This is  obtained using once more Lemma~\ref{lem:well-label-size}, after
  fixing one action per active process. %
  We require from the final
  states that for every %
  active process $p$ the list  contains
  some action that witnesses it.
  In addition, it is required that the first and the last entry of the
  list are on the same process.
\end{itemize}
One can see that it is not necessary to store the content of the message when constructing
the "well-labelings". So by taking the product of these automata, we obtain an automaton
with $|\Proc|^{O(|\Proc|^3)}$ states.

Finally we take the product of all automata $\Bb_p$ such that $p$ is "active" 
and the automaton $\Bb_{all}$.
The resulting automaton has $|\Proc|^{O(|\Proc|^3)}$ states and
verifies the following property:
for every "$\mb$-exchange" $u$, it holds that $u$ is "atomic" iff $\ms(u)$ is accepted by the automaton.
By taking the product 
of this last automaton with the automaton verifying that the "ms-sequence" corresponds to an 
"$\mb$-exchange" (see Lemma~\ref{lem:aut-for-exchange}), we immediately get:%

\begin{mylemma}[atomic-exchange]
  Let $\Aa$ be a "CFM", $g, g'$ two "global states" of $\Aa$
  and $D, D' \subseteq \Proc$.
One can construct an automaton $\Bb$ with 
$O(|G|^3\cdot
|\Proc|^{O(|\Proc|^3)})$ states, such that
\[
  L(\Bb) = \set{v \in (S \cup \bar{S})^* \mid \exists u \text{ "atomic"
      "$\mb$-exchange" } \text{ s.t. } v=\ms(u) \text{ and } (g,D)
    \executable{u}_{\Aa} (g',D')}
\]
\end{mylemma}

\subsection{Verifying $\mb$-synchronizability}\label{sec:algorithm}

To check  non "$\mb$-synchronizability" we look for an "$\mb$-viable" "trace" that is
not equivalent to a $\astm$-product of "$\mb$-exchanges". Such a  \emph{witness}
$u$ must contain some "atomic"
factor $v$ that is not equivalent to an "$\mb$-exchange".
In other words, 
$u \equiv u' \astm v \astm u''$ for some $u',u''$, with $v' \notin S^*R^*$ for
every $v \equiv v'$.
It is enough to reason on "atomic" factors, since for any "exchange" $u$
where  $u \equiv u_1 \astm \dots \astm u_n$ with each $u_i$ "atomic",
all factors $u_i$ are also "exchanges".
Note that an "atomic" $v$ is not equivalent to an "$\mb$-exchange" iff
some process in $v$ does a send after a receive.

The next lemmas refer to the structure of \emph{minimal
  witnesses} for non-"$\mb$-synchronizability".

\begin{mylemma}[dec-atomic-receive]
  Let $u=v \,r$ be an "$\mb$-viable" sequence with $r \in R$.
  There exist "$\mb$-atomic" non-empty sequences $v_1,\dots, v_n$ and
  indices $1\le i <j\le n$ such
  that (1) $v \equiv v_1 \astm \cdots \astm v_n$, and (2)  $u
  \equiv v_1 \astm \cdots \astm v_{i-1} \astm w \astm v_{j+1} \astm \cdots
  \astm v_n$ with  $w=(v_i \astm \cdots \astm v_j) \, r$ being
  "$\mb$-atomic". 
\end{mylemma}
\begin{proof}
  The proof follows  by analyzing the additional edges 
  of $\G_\mb(u)$ compared to $\G_\mb(v)$.
  Let $q$ be the process doing $r$.
  The graph $\G_\mb(u)$ is obtained from $\G_\mb(v)$ by adding the
  double edge between $r$ and its matching send $s$, as well as
  edges from $s$ to all "unmatched" sends $s'$ to process $r$.
  Each SCC of $\G_\mb(u)$ is either an SCC of $\G_\mb(v)$, or a union of
  SCCs of $\G_\mb(v)$ and  contains $s,r$.
  If two  SCCs of $\G_\mb(v)$ are included in the same  SCC $C$ of
  $\G_\mb(u)$ and are ordered in $\G_\mb(v)$, then every SCC
  between them is also included in $C$ as well.
\end{proof}

\begin{mylemma}[atomic-dag]
  Let $u=v \, r$ be an "$\mb$-viable" sequence with $r\in R$, such that
  $v$ is not "$\mb$-atomic".
  We denote by $s$ the send event "matched" with $r$ in $u$, and by $q$
  the process of $r$. Then $u$ is "$\mb$-atomic" iff for every
  decomposition $v\equiv v_1\astm \cdots \astm v_n$ with $v_i$
  "$\mb$-atomic" for all $i$: \\ (1) $v_1$ contains $s$ or some
  "unmatched" send to process $q$, and (2) $v_n$ contains $s$ or some
  action performed by process $q$. 
\end{mylemma}%
An example of such a decomposition is shown in Figure~\ref{fig:lemma-3-20}.
\begin{proof}
  First recall from Lemma~\ref{lem:dec-into-atomic} that the decomposition 
  $v\equiv v_1\astm\cdots\astm v_n$ is unique, up to permuting adjacents
  factors that are unordered w.r.t.~$\preceq_\mb^v$.
  By removing
  $r$ from $u$, we  remove some arcs from the "communication graph"
  $H_\mb(u)$ of $u$,
  namely the double
  arc between $s$ and $r$, and the arcs from $s$ to every "unmatched"
  send to $q$.

  For the left-to-right implication we assume that 
  $u$ is "$\mb$-atomic", or equivalently, $H_\mb(u)$ is
  strongly connected by Lemma~\ref{lem:indivisibility-from-inter}.
  For the first point, if $v_1$ has neither $s$ nor an "unmatched" send to $q$, 
  then there would be no back arc  from $v_2 \dots v_n r$ to
  $v_1$ in $H_\mb(u)$. Hence $H_\mb(u)$ would not be strongly connected.
  For the second point, if  $v_n$ does not
  contain $s$,  nor any action on $q$, then it could
  be reordered after $r$ in $u$, and there would be no back arc from $v_n$
  to  $v_1 \dots v_{n-1} r$.
  Again, $H_\mb(u)$ would not be strongly connected.
  In both cases we get a contradiction.

  For the right-to-left implication let $\ell$ be such that $v_\ell$
  contains $s$.
  By the first condition, every $\preceq_\mb^v$-minimal $v_i$ is such that
  in $H_\mb(u)$ there is either a back arc from $v_\ell$ to $v_i$, or
  from $r$ to $v_i$ (if $i=\ell$).
  By the second condition, every $\preceq_\mb^v$-maximal $v_i$ has a
  forward arc to $r$ in $H_\mb(u)$.
  Since there is also a back arc from $r$ to $v_\ell$ and since every
  $v_i$ by itself has a strongly connected "communication graph", we
  get that $H_\mb(u)$ is strongly connected, so $u$ is
  "$\mb$-atomic".   
\end{proof}

\begin{figure}
  \centering
  \begin{subfigure}[c]{0.45\textwidth}
    \scalebox{0.7}{\input{figures/lemma_3_20_msc.tex}}
  \end{subfigure}
  \hspace{0.1\textwidth}
  \begin{subfigure}[c]{0.35\textwidth}
    \vspace{10pt}
    \scalebox{0.7}{\begin{tikzpicture}[remember picture,thick,every node/.style={  circle,
                                                                inner sep = 1.9pt,
                                                                outer sep = 0pt},font=\Large]
    \node[draw] (n1) at (-2,0) {$u_1$};
    \node[draw] (n2) at (0,0) {$u_2$};
    \node[draw] (n3) at (0,-1.5) {\color{red}$u_4$};
    \node[draw] (n4) at (2,0) {$u_3$};
    \node[draw] (n5) at (3.5,-1.5) {$u_5$};
    \node[draw,red,inner sep = 0.8pt,dashed,outer sep = 2pt] (r) at (1.75,-1.5) {$r$};

    \path[->,every node/.style={circle,inner sep = 0pt,outer sep = -3 pt,sloped,above}]
                (n1) edge node {$<_\Proc$} (n2)
                (n2) edge node {$<_\Proc$} (n3)
                (n2) edge node {$<_\Proc$} (n4)
                (n4) edge node {$<_\mb$} (n5)
                (n5) edge node {$<_\Proc$} (r)
    ;
     \path[->, every node/.style={circle,inner sep = 0pt,outer sep = -3 pt,sloped,above}, blue]
         (n3) edge[<->] node[below] {\textit{msg}} (r)
         (n3) edge[bend left=10,dashed] node {$<_\mb$} (n1)
     ;
\end{tikzpicture}}
  \end{subfigure}
  \caption{The "MSC" of an "atomic" sequence. %
  It is not "\mb-synchronizable" by Lemma~\ref{lem:min-witness},
  each $u_i$ consists of message $i$, the indices are $(1,2,3,5)$, and $m=2$.}
  \label{fig:lemma-3-20}
\end{figure}

\begin{mylemma}[min-witness]
  Let $u=v \,r$ be  "$\mb$-viable" with  $r \in R$ and $v$ is "$\mb$-synchronizable".
  Let also $s$ be the send matching $r$ in $u$, and $q$ the process doing
  $r$.
Then $u$ is not "$\mb$-synchronizable" iff
 there exist $(v_i)_{i=1}^n$ with $v \equiv v_1 \ast \dots \ast v_n$,
 indices $1\le i_1 <\dots < i_k \le n$, and $p \in \Proc$  s.t.:
  \begin{enumerate}
  \item Each $v_i$ is  "$\mb$-atomic".
    \item For every $1 \le j <k$ we have $i_j \prec^v_\mb i_{j+1}$.
      \item $v_{i_1}$  contains $s$ or some "unmatched" send to
        process $q$; 
        $v_{i_k}$ contains $s$ or
        some action performed by process $q$.
      \item There exists $1 \le m < k$ such that $v_{i_m}$ contains a receive by $p$ and
        $v_{i_{m+1}}$  a send by $p$.
  \end{enumerate}
\end{mylemma}
\begin{proof}
  First the left-to-right direction.
  By assumption, $u=v\, r$ with $u$ not "$\mb$-synchronizable" and 
  $v$ is "$\mb$-synchronizable".
 
 By Lemma~\ref{lem:dec-atomic-receive} we can assume w.l.o.g.~that $v
 \equiv v_1 \astm \cdots \astm v_n$ and $u
 \equiv v_1 \astm \cdots \astm v_{i-1} \astm w \astm v_{j+1} \astm
 \cdots \astm v_n$, with $w=(v_i \astm \cdots \astm v_j) \, r$ being
 "$\mb$-atomic" (and containing also $s$).

Since $u$ is not "$\mb$-synchronizable", we have that $w$ is not an
"$\mb$-exchange" (since each $v_k$ is an "$\mb$-exchange").
We have $w = (v_i \astm \cdots \astm v_j) \, r$, so there exist some $i
\le \ell < \ell' \le j$ such that $v_\ell$ contains some receive by some
process $p$, and $v_{\ell'}$ some send by $p$.

We  can assume w.l.o.g.~that $i \preceq^w_\mb \ell$ and $\ell'
\preceq^w_\mb j$ in $\skel(w)$.
Note also that $\ell \prec^w_\mb \ell'$ because $v_\ell, v_{\ell'}$ both contain
actions of $p$.
So we get a subsequence of indices $i=i_1< \dots < i_k=j$ satisfying
items (2) and (5) of the statement.
Item (3)  follows from Lemma~\ref{lem:atomic-dag} applied to $w$.

\smallskip

The right-to-left direction is easily checked: because of items
(1)-(4) all events of the sequences $v_{i_1},\dots, v_{i_k}$ are in
the same SCC of $\G_\mb(u)$.
Item (5) says that this SCC cannot be part of an
"$\mb$-exchange". 
\end{proof}

Note that while we can guess "$\mb$-synchronous"
sequences without storing messages (Lemma~\ref{lem:aut-for-exchange}),
we need to be careful when guessing $u$
in Lemma~\ref{lem:min-witness} so that it is "$\mb$-viable".
For instance, by reversing message 2 in Figure~\ref{fig:lemma-3-20}
the sequence becomes non-"\mb-viable". %

The next lemma shows how to check the existence of a $(\lhb \cup
<_\mb)$-path between two positions of an "ms-sequence", using
the automaton from Lemma~\ref{lem:causality-check-auto-marked}.

\begin{mylemma}[causality-check-auto]
  One can construct  an automaton $\Dd$ with $O(|\Proc|)$ states 
  over the alphabet $(S \cup \bS) \times \set{\circ,\bullet} \cup \set{\#}$
  with the following properties: 
  \begin{enumerate}
  \item $\Dd$ accepts only words from $(\S^* \#)^*$ containing exactly
    two positions in $(S \cup \bS) \times \set{\bullet}$.
\item For every $u=u_1 \astm \dots \astm
  u_n$ "\mb-viable", with each $u_i$ an "exchange",
  $\Dd$ accepts tagged $v=\ms(u_1) \# \dots \# \ms(u_n)$ iff,
  there is a $(\lhb \cup <_\mb)$-path from $u[i]$ to $u[j]$, where 
  $i<j$ are the positions of $u$ corresponding to the positions tagged by $\bullet$ in $v$.
  \end{enumerate}
\end{mylemma}

\begin{proof}
  As we work on "ms-sequences" and not on "marked" sequences 
  as in Lemma~\ref{lem:causality-check-auto-marked}, we need to add $\#$ at the
  end of our "exchanges". We will let the automaton guess when an "exchange" ends. 
  Sometimes it is trivial, as it just needs to see a receive action. 
  However, if no "matched" send had been seen since the last $\#$, the automaton can
  split each "unmatched" send in its own "exchange", or group them together.
  In the end the automata will read $\ms(u_1) \# \dots \# \ms(u_n)$ for
  the partition $u_1 \astm \dots \astm u_n$ it has guessed.

  We then change the third rule from Lemma~\ref{lem:causality-check-auto-marked}, so that
  the automaton goes from $(\msg,S,p)$ to $(\msg,R,p)$ when reading a $\#$.
\end{proof}

We have now all ingredients to show our main result.
We use Lemma~\ref{lem:min-witness} to guess the witness sequence, "exchange" by "exchange",
and to be sure that the sequence is "\mb-viable"
we rely on Lemmas~\ref{lem:mb-viable} and
\ref{lem:causality-check-auto}, complementing the automaton on-the-fly.
The lower bound is obtained, as before,
by reduction from the intersection emptiness problem for finite automata.

\begin{mytheorem}[mb-synchronizability]
  The question whether a "CFM" is "$\mb$-synchronizable" is \PSPACE-complete.
\end{mytheorem}

\begin{proof}
  For the upper bound we look for a witness execution $u$ which is
  not "$\mb$-synchronizable" and of minimal length.
  Hence $u=v\, a$ is an "$\mb$-viable" sequence such that $v$ is
  "$\mb$-synchronizable". Note that the last action $a$ is a receive $r$, since otherwise $u$
  would be "$\mb$-synchronizable", too. 
  
  We use Lemma~\ref{lem:min-witness} to guess such a
  minimal witness $u = v\, r$.
  Recall that $q$ is the process executing $r$, and $s$ the matching
  send of $r$ in $u$.

First we rely on the automaton of Lemma~\ref{lem:atomic-exchange} in
order to guess the "atomic" "exchanges" $v_i$ composing $v$ on-the-fly.
At the same time we guess the subsequence of indices $i_1< \cdots <
i_k$ and the events that witness 
that $i_j \prec^v_\mb i_{j+1}$
(\cf~Definition~\ref{def:order-atoms}).

  We keep record of the current pair $(g,D)$, where $g$ is a global
  state of the "CFM" and $D$ a set of "deaf" processes,  as we guess each $v_i$,
  to check that the sequence $v$ labels an "execution".
  When we process $v_{i_k}$, we remember its alphabet over
  $S\cup\bar{S}$ until we guess $v_{i_{k+1}}$, 
  and check that $i_k \prec^v_\mb i_{k+1}$ (\cf~Remark~\ref{rem:order-on-sequence}).
  We also guess  $m$ as of item (4) in Lemma~\ref{lem:min-witness},
  and check the condition.
  After we have done $v_n$, we must have reached $(g,D)$ such that the receive
  $r$ can be done in state $g$.
  By verifying that $u$ is equivalent to an "$\mb$-viable" sequence as described below, we know
  that  $s$ is "matched" with $r$.

  We check finally that $v_1 \ast \dots \ast v_n \, r$ is equivalent to an
  "$\mb$-viable" sequence (this suffices by Remark~\ref{rem:viable-vs-similar}). %
  From Lemma~\ref{lem:mb-viable} 
  we know that $u$ is equivalent to an "$\mb$-viable" sequence iff there is no "unmatched" send $s'$ to $q$
  s.t.~there is a $(\lehb \cup \mbord)$-path from $s'$ to  $s$ in
  $v$.
  For this we use the complement $\Dd^{co}$  of the automaton $\Dd$ of Lemma~\ref{lem:causality-check-auto}
  which is exponential in $|\Proc|$ but can be constructed
  on-the-fly in linear space.
  We make one copy $\Dd^{co}(p')$ of $\Dd^{co}$ for every process
  $p'\not= q$.
  Each $\Dd^{co}(p')$ \emph{tags} the first "unmatched"
  send of type $p'!q$ and $s$ with $\bullet$.
  We make every $\Dd^{co}(p')$  read the tagged $\ms(v_1) \# \dots \# \ms(v_n)$
  by adding the $\#$ after each "atomic" "$\mb$-exchange" we read.
  Each $\Dd^{co}(p')$
  should accept.
 This guarantees that no send of type $\overline{p'!q}$ has a
 $(\lehb \cup \mbord)$-path to $s$.

  For the lower bound, we use the same reduction as in Theorem~\ref{thm:reach-for-mb-sync},
  and if we reach $(\texttt{accept})_{p\in\Proc}$, we use two other processes to do a
  non-"\mb-synchronizable" gadget (see
  Figure~\ref{fig:non-sync-gadget}).
  This way, the "CFM" is "$\mb$-synchronizable" if and only
  if the intersection of the automata $\Aa_1,\dots,\Aa_n$ is empty.
\end{proof}

Theorem~\ref{thm:mb-synchronizability} yields two interesting
corollaries.
We define a \emph{$k$-\mb-exchange} as any "$\mb$-viable" sequence in $S^{\leq k}R^*$.
An "$\mb$-viable" sequence is \emph{$k$-\mb-synchronous} if it is a
$\astm$-product of $k$-\mb-exchanges,
and is called \emph{$k$-\mb-synchronizable} if it is equivalent to a $k$-\mb-synchronous sequence.
A "CFM" $\Aa$ is \emph{$k$-$\mb$-synchronizable}
if every "trace" in $\Ex_\mb(\Aa)$ is $k$-\mb-synchronizable.
The next result has been shown decidable in
\cite{BolligGFLLS21Concur} %
(with non-elementary complexity): %
\begin{mytheorem}[k-sync]
  Let $k$ be an integer given in binary. The question whether a  "CFM" is
  $k$-$\mb$-synchronizable is \PSPACE-complete.
  The lower bound already holds for $k$ in unary.
\end{mytheorem}

\begin{proof}
  Using Theorem~\ref{thm:mb-synchronizability} we first check that the
  "CFM" is "$\mb$-synchronizable".
  Then we  use the automaton $\Cc$ from
  Lemma~\ref{lem:aut-for-exchange} to compute pairs $(g,D)$ of global
  state and set of "deaf" processes that are reachable by some
  "$\mb$-synchronous" sequence.
  Finally we check whether the automaton of
  Lemma~\ref{lem:atomic-exchange} accepts only "exchanges" of size at
  most $k$.
  Since the size of our automata is exponential the test can be done
  in \PSPACE.
  The lower bound can be obtained as in the proof of
  Theorem~\ref{thm:mb-synchronizability}  (see 
  Figure~\ref{fig:inter-auto-msc}).
\end{proof}

For the second result and weak
synchronizability, decidability was obtained in~\cite{GiustoLL23}.
Our proof based on automata seems more direct and simpler than the one of~\cite{GiustoLL23}:

\begin{mytheorem}[guess-bound]
  The question whether for a given "CFM" $\Aa$ there exists some $k$ such
  that $\Aa$ is $k$-$\mb$-synchronizable, is \PSPACE-complete. %
\end{mytheorem}

\begin{proof}
  For the upper bound we proceed as in the previous proof.
  The difference is that at the end we check whether the automaton of
  Lemma~\ref{lem:atomic-exchange} 
  accepts an infinite language from a reachable pair $(g,D)$.
  The language of this automaton is infinite iff there is no $k$ as
  stated by the theorem.
  The lower bound can be obtained as in the proof of
  Theorem~\ref{thm:mb-synchronizability}.
\end{proof}

\section{Other notions of synchronous systems}
\label{sec:comp}
\begin{figure}[t]
    \centering
    \scalebox{0.7}{\input{figures/msc_non_sync.tex}}\\
    {\footnotesize$p_2!p_1(b)\,p_2!p_3(c)\,p_3?p_2(c)\,p_3!p_2(d)\,p_2?p_3(d)\,
    p_1!p_2(a)\,p_1?p_2(b)$}
    \caption{A \emph{weakly-synchronous}
      sequence~\cite{BolligGFLLS21Concur} that is not
    "$\mb$-synchronizable".}
    \label{fig:msc-non-sync}
\end{figure}

The notion of synchronizability used in this paper is not the only one present
in the literature.
In this section, we compare it with some of the alternative notions.

\subsection{Weak synchronizability}

\emph{Weak $k$-synchronizability} was introduced
in~\cite{bouajjani2018completeness}, where it is simply called
\emph{$k$-synchronizability}. This notion was extended
in~\cite{BolligGFLLS21Concur,GiustoFLL23,GiustoLL23} by removing the bound on the
size of the exchanges, yielding \emph{weak
synchronizability}~\cite{BolligGFLLS21Concur}.

The main difference between the notion of synchronizability used in~\cite{bouajjani2018completeness,BolligGFLLS21Concur,GiustoFLL23,GiustoLL23}
and our definition
is related to the notion of equivalence between action sequences. In this paper,
a (mailbox) sequence is called \emph{synchronizable} (more precisely, \mb-synchronizable)
if it is equivalent to a $\astm$-product of \mb-exchanges. In contrast, a
(mailbox) sequence is \emph{weakly synchronizable} if it is equivalent to a 
$\astp$-product of \mb-exchanges.  Note that using the $\astp$-product
does not guarantee that the sequence is \mb-viable. 
As a consequence, weak-synchronizability yields more
synchronizable sequences, however some  can be spurious (this leads to  the
notion of \emph{strong synchronizability} in \cite{BolligGFLLS21Concur}, which
is the same as  synchronizability here).
  In particular one cannot use the decomposition into "exchanges" 
  from~\cite{bouajjani2018completeness,BolligGFLLS21Concur} to
check regular properties of "executions",  
as we do in Section~\ref{sec:properties}.

Figure~\ref{fig:msc-non-sync} shows
an example distinguishing the definitions.
The sequence $p_1!p_2(a) \, \astp\,
     p_2 ! p_1 (b)\, p_1 ? p_2 (b)\, \astp \,p_2 !p_3 (c)\,  p_3 ? p_2(c)\,\astp\,p_3 !p_2 (d)\,  p_2 ? p_3(d)$ corresponds to a decomposition in
"exchanges" according
to~\cite{bouajjani2018completeness,BolligGFLLS21Concur}, but it is not
"$\mb$-viable".
Note that in the case all messages
are "matched", the two products are equivalent.

\subsection{Send-synchronizability}
This notion was proposed to address problems about message choreography. In \emph{message choreography}, 
the system is checked against regular properties on sequences of messages instead of sequences
of actions as here.

\emph{Send-synchronizability} was first introduced in~\cite{FuBS04}, where a system is called \emph{synchronizable}
if its asynchronous executions that reach a final state have the same set of
projections on send actions as
the executions where each send is immediately followed by its matching receive.
This notion was used in several works 
afterwards~\cite{BB11,BASU201660,FinkelL23,di2024synchronisability},
modulo some variants (no condition on final state, same state reached by 
the synchronous and asynchronous execution, send actions do or do not need to be
matched, etc \dots).

We recall the definition given in~\cite{BASU201660} in the terminology used in
this work.
Given a "process network" $\Nn$,
a "CFM" $\Aa$ is called \emph{$\Nn$-send-synchronizable} if,
for every trace $u$ of $\Aa$ over $\Nn$,
there is a fully matched $1$-$\Nn$-synchronous trace $v$ of $\Aa$ over $\Nn$
such that $u$ and $v$ have the same projection on sends.
This notion was later studied in~\cite{FinkelL23} to show
that it is undecidable to know if a "CFM" is
\ptp-send-synchronizable.
For now, we do not know if undecidability holds also for \mb-send-synchronizability~\cite{di2024synchronisability}.

Although they appear to be similar, \mb-send-synchronizability and
$1$-\mb-synchronizability are incomparable.
For example, the "CFM" whose executions are given by the MSC of 
Figure~\ref{fig:onesync_not_l_sync} has all its traces $1$-\mb-synchronizable,
but it is not \mb-send-synchronizable because of the (asynchronous) trace
$p_3!p_2(b)\,p_2!p_1(a)\,p_2?p_3(b)\,p_2?p_2(a)$, the projection of which
does not correspond to any $1$-\mb-synchronous trace.

On the other hand, the "CFM" in Figure~\ref{fig:l_sync_not_onesync} is \mb-send-synchronizable, but it is not 
$1$-\mb-synchronizable. This example admits the trace $p!q(a)\,q!p(b)\,q?p(a)\,p?q(b)$,
which is not equivalent to any $1$-\mb-synchronous sequence.
However,
every word in $\{p!q(a), q!p(b)\}^*$ is the projection on sends of
a fully matched $1$-\mb-synchronous trace,
so this "CFM" is \mb-send-synchronizable.

\begin{figure}[t]
    \centering
    \begin{minipage}{0.49\textwidth}
        \centering
        \scalebox{0.8}{%
            \begin{msc}[font=\Large]{}
                \declinst{p0}{}{$p_1$}
                \declinst{p1}{}{$p_2$}
                \declinst{p2}{}{$p_3$}
                \mess{$a$}{p1}{p0}
                \nextlevel
                \mess{$b$}{p2}{p1}
                \nextlevel
            \end{msc}%
        }
        \caption{A $1$-\mb-synchronizable MSC that is not \mb-send-synchronizable}
        \label{fig:onesync_not_l_sync}
    \end{minipage}
    \begin{minipage}{.49\textwidth}
        \centering
        \begin{tikzpicture}[auto, every state/.style={minimum size=15pt}]
            \node        at(-0.7,0) {\textbf{p:}};
            \node[state] (s) {};
            \path[->]   (s) edge [loop above] node {$p!q(a)$} ()
                        (s) edge [loop below] node {$p?q(b)$} ();
            \begin{scope}[xshift=75pt]
                \node        at(-0.7,0) {\textbf{q:}};
                \node[state] (s) {};
                \path[->]   (s) edge [loop above] node {$q!p(b)$} ()
                            (s) edge [loop below] node {$q?p(a)$} ();
            \end{scope}
        \end{tikzpicture}
        \caption{A \mb-send-synchronizable "CFM" that is not $1$-\mb-synchronizable}
        \label{fig:l_sync_not_onesync}
    \end{minipage}
\end{figure}

\section{Conclusion}
We have introduced a novel automata-based approach to reason about
communication in the $\texttt{sr}$-round mailbox model.
We showed that knowing whether a system complies with this model is
\PSPACE-complete.
An interesting theoretical question is whether we can apply similar techniques to
other types of communication.
On the practical side it would be interesting to implement our
algorithms and compare \eg~with existing tools like
Soter~\cite{DOsualdoKO13} that targets safety properties for
a relaxed model of Erlang.
Our automata-based techniques may be easier to implement than 
previous approaches, and could even adapt to a dynamic setting.

\bibliographystyle{alphaurl}
\bibliography{biblio.bib}

\end{document}